# Broad-band spectroscopy of a vanadyl porphyrin: a model electronuclear spin qudit

Ignacio Gimeno,[a] Ainhoa Urtizberea,[a,b] Juan Román-Roche,[a] David Zueco,[a] Agustín Camón,[a] Pablo J. Alonso,[a] Olivier Roubeau,*[a] and Fernando Luis[*a]

[a] Instituto de Nanociencia y Materiales de Aragón, CSIC and Universidad de Zaragoza, 50009 Zaragoza, Spain. E-mail: roubeau@unizar.es; fluis@unizar.es
[b] Centro Universitario de la Defensa, Carretera de Huesca s/n, 50090 Zaragoza, Spain.

We explore how to encode more than a qubit in vanadyl porphyrin molecules hosting a $S = ½$ electronic spin coupled to a $I = 7/2$ nuclear spin. The spin Hamiltonian and its parameters, as well as the spin dynamics, have been determined via a combination of electron paramagnetic resonance, heat capacity, magnetization and on-chip magnetic spectroscopy experiments performed on single crystals. We find low temperature spin coherence times of micro-seconds and spin relaxation times longer than a second. For sufficiently strong magnetic fields ($B > 0.1$ T, corresponding to resonance frequencies of 9-10 GHz) these properties make vanadyl porphyrin molecules suitable qubit realizations. The presence of multiple equispaced nuclear spin levels then merely provides 8 alternatives to define the '1' and '0' basis states. For lower magnetic fields ($B < 0.1$ T), and lower frequencies ($< 2$ GHz), we find spectroscopic signatures of a sizeable electronuclear entanglement. This effect generates a larger set of allowed transitions between different electronuclear spin states and removes their degeneracies. Under these conditions, we show that each molecule fulfills the conditions to act as a universal 4-qubit processor or, equivalently, as a $d = 16$ qudit. These findings widen the catalogue of chemically designed systems able to implement non-trivial quantum functionalities, such as quantum simulations and, especially, quantum error correction at the molecular level.

**Introduction**

Magnetic molecules are appealing candidates to form the building blocks of future quantum technologies.[1] One of their characteristic traits is the ability to tune their properties by chemical design, e.g. to enhance the spin coherence times[2-4] or to gain a sufficiently fast control of their quantum states by means of external stimuli, such as microwave magnetic[5] or electric[6] fields and even visible light.[7] Another remarkable feature is the possibility of hosting multiple qubits, or $d$-dimensional qudits, in a single molecule.[8] This possibility allows integrating at the molecular level non-trivial functionalities, like quantum error correction[9] or simple quantum codes and simulations,[10] and constitutes a potential competitive advantage over other solid-state quantum platforms.

A relatively straightforward strategy to embody multiple qubits is to synthesize molecules having several, weakly coupled, magnetic metal centres.[8a,11] As a somewhat simpler alternative, one can also use internal degrees of freedom, arising from electronic spin projections of $S > \frac{1}{2}$ metal ions[12] or from nuclear spin states.[8b,9b,13] One advantage of this second approach is that it facilitates the synthesis of magnetically diluted single crystals, which can combine sufficiently long coherence times with the ability of addressing specific quantum operations. In addition, these mononuclear complexes are often more robust, thus easier to interface with solid state devices and circuits.[6a,10a,14] The choice of suitable candidates is however quite limited by the need of ensuring a proper addressability of the relevant transitions linking different levels of each qudit. Among electronic spins, the magnetic anisotropy then needs to be very weak, in order to keep resonant frequencies accessible (typically meaning within 1 to 10 GHz). In practice, this has restricted the search to Gd based qudits.[12] In the case of nuclear spin qudits, the same condition imposes a sufficiently strong quadrupolar interaction. The latter criterion then seems to exclude several a priori promising systems, such as the family of vanadyl based molecules that includes among its members some of the rare examples of quantum spin coherence surviving up to room temperature[15] and record spin coherence times close to milliseconds.[16]

In this work, we show that the sought level anharmonicity required to define multiple qubits, or a qudit, using electronuclear spin states can be achieved even when the quadrupolar interaction is negligible. A broad band spectroscopy study of vanadyl porphyrin crystals provides direct and detailed information on the magnetic energy level structure. Extending these measurements to very low temperatures allows exploring low frequencies and magnetic fields. The entanglement of the vanadyl $S = 1/2$ electronic spin with its $I = 7/2$ nuclear spin results then in a series of unequally spaced levels. In this regime, transition rates between different nuclear spin states become also comparable to those of electronic transitions. On basis of this study, we discuss the possibility of using vanadyl molecules to perform universal quantum operations, either as 4-qubit processors or as qudits with dimension $d = 16$.

**Results and discussion**

**Synthesis, structure and spin Hamiltonian of [VOTCPPEt]**
For the present study, we focus on the vanadyl porphyrin [VO(TCPPEt)] (**1$^{VO}$**, H$_2$TCPPEt = 5,10,15,20-tetrakis(4-carboxyphenyl)porphine tetraethyl ester), a molecule we previously used as precursor to form a 2D framework.[17] Although the structure and spin dynamics of **1**$^{VO}$ were not determined, the vanadyl porphyrin node in the 2D framework did show some reasonably long quantum coherence. Besides, **1**$^{VO}$ is readily obtained by a simple reaction of vanadyl sulphate and the free-base H$_6$TCPP in ethanol under solvothermal conditions, and most importantly in the form of robust large single-crystals. Interestingly, crystals of the titanyl analogue [TiO(TCPPEt)] (**1**$^{TiO}$) can be obtained under similar synthetic conditions, *albeit* only with partial metallation, even using a large excess of the titanyl starting salt. The free-base ligand H$_2$TCPPEt, obtained under the same conditions, and both vanadyl and titanyl complexes were found by single-crystal diffraction to be isostructural, which in turn allowed to isolate crystals containing the vanadyl porphyrin diluted in a diamagnetic host, **1**$^{VO}_{3\%}$ (see ESI for details).

All three materials H$_2$TCPPEt, **1**$^{VO}$ and **1**$^{TiO}$ crystallize unsolvated in the monoclinic space group *P*2$_1$/n, the asymmetric unit coinciding with half the neutral MTCPPEt molecule (M = H$_2$, VO, TiO). In **1**$^{VO}$, the V$^{4+}$ ion is coordinated equatorially by the four pyrrole N (M-N bond length in the range 1.994(4)-2.172(4) Å), but lies slightly out of the N$_4$ porphyrin plane, at 0.491 Å (Fig. 1a and b).

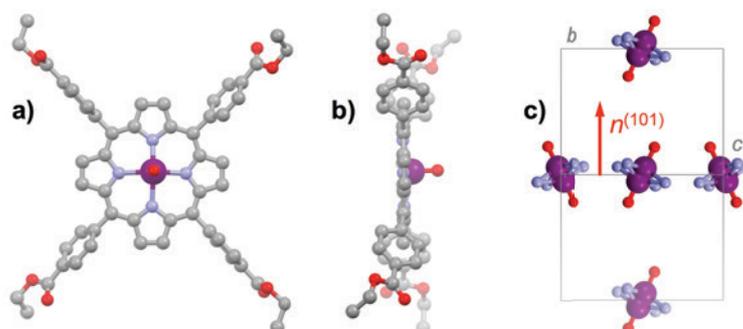

**Figure 1.** a) Top and b) side views of the [VOTCPPEt] molecule in the structure of **1**$^{VO}$. Colour code: plum, V; red, O; light blue, N; grey, C. Hydrogens are omitted for clarity. c) View along the 101 plane of one unit cell of **1**$^{VO}$, showing the two types of molecules and the disorder in the V=O moiety orientation. The average V=O orientation corresponds to the normal to the 101 plane, shown as a red arrow. Only the VON$_4$ coordination spheres are shown for clarity.

The axial V=O bond (1.556(9) Å) is thus disordered over two equivalent orientations by symmetry, due to the inversion center close to the center of the porphyrin N4 square. The porphyrin ring is basically flat with no significant distortion (Fig. 1b), the largest deviation being for the external pyrrole carbons, albeit in all cases at less than 0.102 Å from the N$_4$ porphyrin plane. One of the two unique ethyl groups is also disordered over two positions. The molecules have two orientations in the cell as a result of the 2-fold screw axis. The average VO vector is however normal to the 101 plane, since this plane is the bisector of the N$_4$ porphyrin plane (Fig. 1c). The molecules with identical orientation form supramolecular 1D chains through C-H···π interactions involving the four pyrrole rings, these chains being connected into planes through π···π interactions between pyrrole ring and COO ester moieties of molecules with different orientations (Figs. S1-S2). The shortest V···V separation corresponds to the translation along the *a* axis, at 9.394 Å, while those within the supramolecular chains and between molecules with different orientations are respectively 10.856 Å and 12.853 Å.

The continuous wave (CW) X-band EPR spectra of a frozen solution of **1**[VO] in toluene:CDCl$_3$ (1:1, 0.46 mmol/L), **1**[VO]$_{sol}$, and of polycrystalline **1**[VO] are shown in Fig. 2 (top traces). We have also studied the rotational CW EPR spectra of single crystals of **1**[VO] and of **1**[VO]$_{3\%}$ (Fig. 2 bottom and Figs. S4, S8 and S9). All are characteristic of an axial vanadyl spin, in agreement with the local symmetry at the porphyrin site. In particular, it appears that the two orientations of the molecules in a single-crystal are magnetically equivalent when the magnetic field is applied perpendicular to the long crystal axis, labelled here as the laboratory Z axis (see Fig. S3). Clearly this must correspond to the situation where the magnetic field is applied in the 101 crystal plane, perpendicular to the average V=O vector (see Fig. 1c). Moreover, there are two perpendicular orientations of the field within the laboratory XY plane, $x_C$ and $y_C$, for which the splitting of the EPR is respectively minimal and maximal. This allows defining the magnetic axes for each molecule, taking advantage of the fact the symmetry of the rotational diagrams is solely reflecting the local molecular symmetry. Taking into account the structural information to define the angle between the two local gyromagnetic z axes, the rotational data is reproduced very satisfactorily using an axial spin-Hamiltonian for each molecule

$$\mathcal{H} = -g_\perp \mu_B (S_x B_x + S_y B_y) - g_\parallel \mu_B S_z B_z \\ + A_\perp (I_x S_x + I_y S_y) + A_\parallel I_z S_z \quad (1)$$

that includes the electronic Zeeman interaction and the hyperfine interaction with the $I$ = 7/2 $^{51}$V nucleus (natural abundance 99.75 %), both referred to the molecular axes $x$, $y$ and $z$, and provides a robust definition of the parameters $g_\parallel$ = 1.963, $g_\perp$ = 1.99, $A_\parallel$ = 475 MHz and $A_\perp$ = 172 MHz. The same model also reproduces the spectra of **1**[VO]$_{sol}$, and polycrystalline **1**[VO] shown in Fig. 2.

The similarity of the CW-EPR spectra for the pure material and its frozen and solid solutions (Fig. 2 and Fig. S10) points at the absence of significant intermolecular magnetic interactions. This is also supported by the equilibrium magnetic susceptibility $c$ of polycrystalline **1**[VO] that follows the Curie law down to 2 K with $C$ = 0.367 cm$^3$mol$^{-1}$, while the magnetization *vs.* field data is well described by the $S$ = ½ Brillouin function for $g$ = 1.98 (Figure S18). Both sets of data are also in excellent agreement with the $g$ tensor derived from CW-EPR.

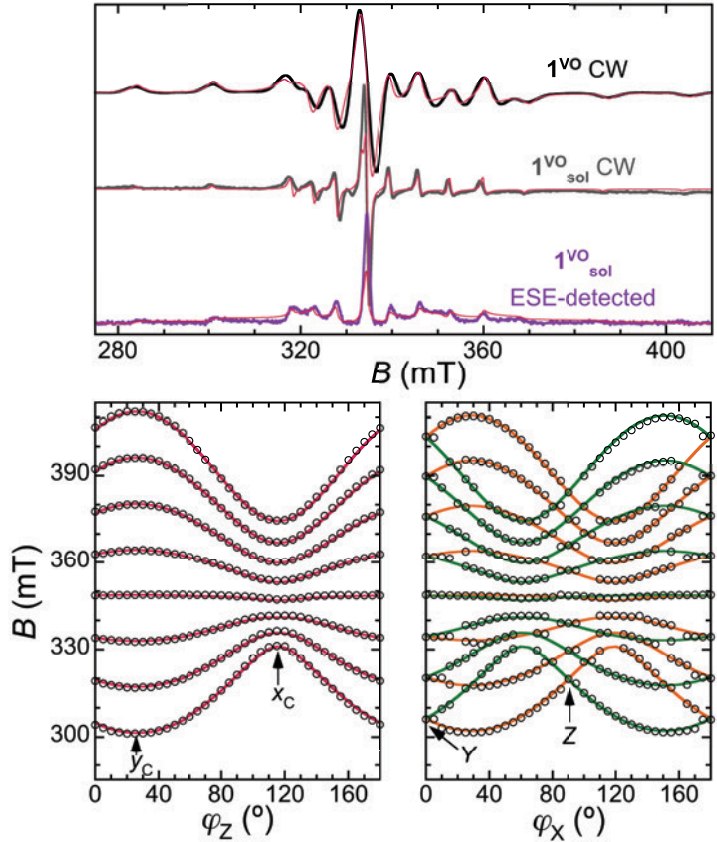

**Figure 2.** *Top*: X-band CW-EPR spectra of **1$^{VO}$** at RT (top trace) and the frozen solution **1$^{VO}_{sol}$** at 30 K (middle trace), and 2p-ESE-detected EPR spectrum of **1$^{VO}_{sol}$** ($\tau$ = 200 ns, 30 K, bottom trace). The thin red lines are simulations using an axial SH (Eq. (1)) with $g_{\parallel}$ = 1.963, $g_{\perp}$ = 1.99, $A_{\parallel}$ = 475 MHz and $A_{\perp}$ = 172 MHz. *Bottom*: Rotational diagrams for a single crystal of **1$^{VO}$** at RT. The empty circles represent the positions of the center of the lines in the CW-EPR spectra upon rotating the crystal around the Z (left) and X (middle) laboratory axes (see also ESI). Full lines are the corresponding positions calculated with the spin Hamiltonian. The contribution from the two magnetically inequivalent orientations of the molecules is represented in different colours (orange and green).

The specific heat $c_P/R$ is also sensitive to the structure of magnetic energy levels. Its temperature dependence for a single-crystal of **1$^{VO}$** in zero-field shows a low temperature tail, which can be associated with the hyperfine splitting of the vanadyl nuclear spin levels (Fig. S24). Under a magnetic field, a Schottky anomaly associated with the Zeeman splitting between electronic spin-up and spin down states is observed, its maximum becoming detectable in the experimental temperature range ($T$ > 0.35 K) for fields above 2 T. At higher temperatures, the specific heat $c_P/R$ becomes dominated by excitations of vibrational modes. The Schottky broad maxima are well accounted for by numerical calculations based on the parameters of the spin Hamiltonian derived above, thus fully supporting these. An exception is observed for intermediate fields of 0.25, 0.5 and 1 T, for which the expected rise in $c_P/R$ associated with the Schottky anomaly is either not detected or reduced with respect to the simulation. As discussed below, this is due to the slow magnetization dynamics of the vanadyl spins (see also Fig. S25).

## [VO(TCPPEt)] as a spin qubit: spin coherence and spin relaxation time scales

Knowledge of the temperature dependence of the spin relaxation timescales, described through the spin-lattice relaxation time $T_1$, is important for various reasons. First, because it represents the ultimate limitation of $T_2$ when all other sources of decoherence are suppressed, i.e. $T_2 \leq T_1$. At higher temperatures, $T_2$ is indeed often limited by the decreasing $T_1$. On the other hand, $1/T_1$ sets the speed at which the spins decay towards the ground state at very low $T$, thus also the initialization process rate in quantum information applications.

We have first studied the spin-lattice relaxation of the VOTCPPEt molecule in the frozen solution $\mathbf{1^{VO}_{sol}}$ through TD EPR at X-band frequencies. Inversion recovery experiments were performed between 6 K and 120 K at $B$ = 346.6 mT, which corresponds to the most intense resonance line in the CW and ESE-detected spectra (see Fig. 2). The experimental dependence on $t_d$, the delay time after the initial inversion pulse, is reproduced by a stretched exponential, written as $\exp\{-(t_d/\beta T_1)^\beta\}$,[18] thus allowing to derive a mean $T_1$ value (Fig. S11 and Table S3). From about 50 ms at 6 K, $T_1$ decreases continuously with increasing temperature, to reach about 20 ms at 120 K (Fig. 3). The temperature dependence can be reproduced considering direct and Raman relaxation processes (see caption of Fig. 3). Although the direct process is dominant at the lowest temperature of 6 K, Raman processes are already effective and then dominate the temperature dependence of $T_1$ at increasing temperatures.

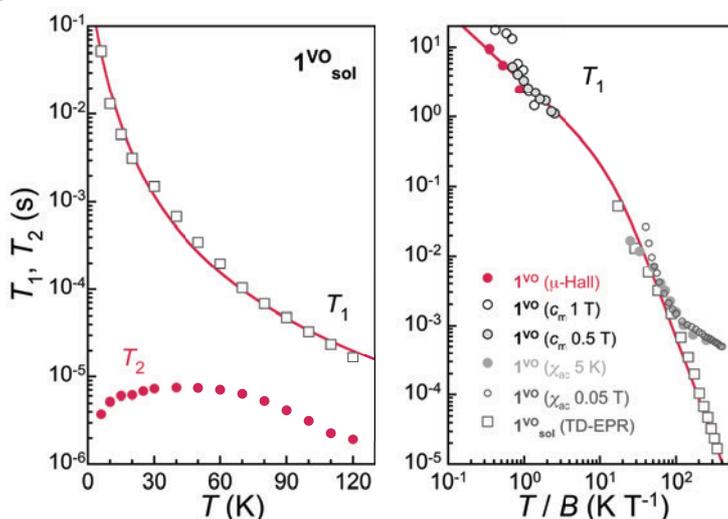

**Figure 3.** *Left*: temperature dependence of the mean longitudinal relaxation time $T_1$ and of the coherence time $T_2$ for $\mathbf{1^{VO}_{sol}}$ at 346.6 mT. The red line is a fit of the data to $1/T_1 = aT + cT^n$, with $a$ = 2.2(2) s$^{-1}$T$^{-2}$K$^{-1}$ and $c$ = 0.029(3) s$^{-1}$K$^{-3}$ and $n$ = 3. *Right*: Comparison of the relaxation times $T_1$ determined by different techniques either on $\mathbf{1^{VO}}$ or its frozen solution $\mathbf{1^{VO}_{sol}}$ as indicated. The red line is the expression $1/T_1 = a'(T/B) + c'(T/B)^3$ with $a'$ = 0.324(9) s$^{-1}$T$^{-3}$K$^{-1}$ and $c'$ = 0.0145 s$^{-1}$T$^3$K$^{-3}$ and allows to describe most data at intermediate fields.

In the solid-state, the magnetization dynamics was studied through *ac* magnetic susceptibility measurements on polycrystalline $\mathbf{1^{VO}}$ (Figs. S19-S21). Upon the application of a *dc* field, an out-of-phase component of the susceptibility $\chi''$ shows up, corresponding to a decrease in the real susceptibility $\chi'$. For spin 1/2 systems, the characteristic relaxation time is a direct measure of $T_1$. It was determined, both at 5 K for increasing *dc* fields and at 500 Oe for *T* ranging from 2 to 20 K, by fitting the isothermal frequency dependence of $\chi'$ and $\chi''$ with a generalized Debye model.

The spin-lattice relaxation time was also estimated from heat capacity measurements on a single-crystal of **1$^{VO}$**. As mentioned above, a sharp drop is observed in the magnetic component $c_m$ measured at certain field values (see e.g. the data measured at *B* = 1 T in Fig. S25). This deviation from equilibrium occurs when phonon-induced relaxation processes become too slow as compared with the time-scale of the relaxation measurement.[19] By comparing measurements done with short (out of equilibrium) and long (equilibrium) experimental times, the characteristic time of the spin thermalization, which is equivalent to $T_1$, can be derived. Finally, m-Hall magnetization measurements performed on a single-crystal of **1$^{VO}$** provide yet another evidence of the slow spin dynamics. For $T \leq 1$ K, $T_1$ becomes so long that the system shows magnetic hysteresis (Fig. S27) despite the absence of an activation barrier for the spin reversal. Dc relaxation measurements performed at fixed temperatures and magnetic fields allow extracting $T_1$.

Values of $T_1$ derived from all measurements are plotted together *vs. T / B* in Fig. 3 right. Overall, there is good agreement among data obtained by different methods, except for those extracted from *ac* susceptibility at low fields. This is understandable as the direct relaxation process becomes then less significant, while it is likely the dominant process at low temperatures and under intermediate to high fields. In fact, and quite remarkably, most data can be described considering direct and Raman processes (see caption of Fig. 3), confirming that the spin relaxation is dominated by direct processes at low temperatures, at least for moderate fields in the range 0.1-1.5 T.

The relaxation times for the [VO(TCPPEt)] molecule are remarkably long, for example about two times longer than those reported recently for dicatechol vanadyl complexes as the highest for vanadyl-based compounds.[20] As discussed previously, this is likely the consequence of the square pyramidal environment of the V(IV) ion and the overall rigidity of its coordination sphere. Both reduce the probability that the electronic spin couples significantly with low-lying vibrational modes. The longer $T_1$ in **1$^{VO}$** with respect to other vanadyl bis-chelates can probably be ascribed to the higher rigidity of the macrocyclic tetra-chelating porphyrin. It is in this respect also to note that the porphyrin seems to outperform its phtalocyanine analogue, as VOPc was reported to have similar but shorter spin-relaxation times than those of **1$^{VO}$**.[15,21] Importantly, such long spin-lattice relaxation times ensure that $T_1$ will not be limiting the coherence time-scale. On the other hand, they can represent a drawback for an efficient qubit initialization.

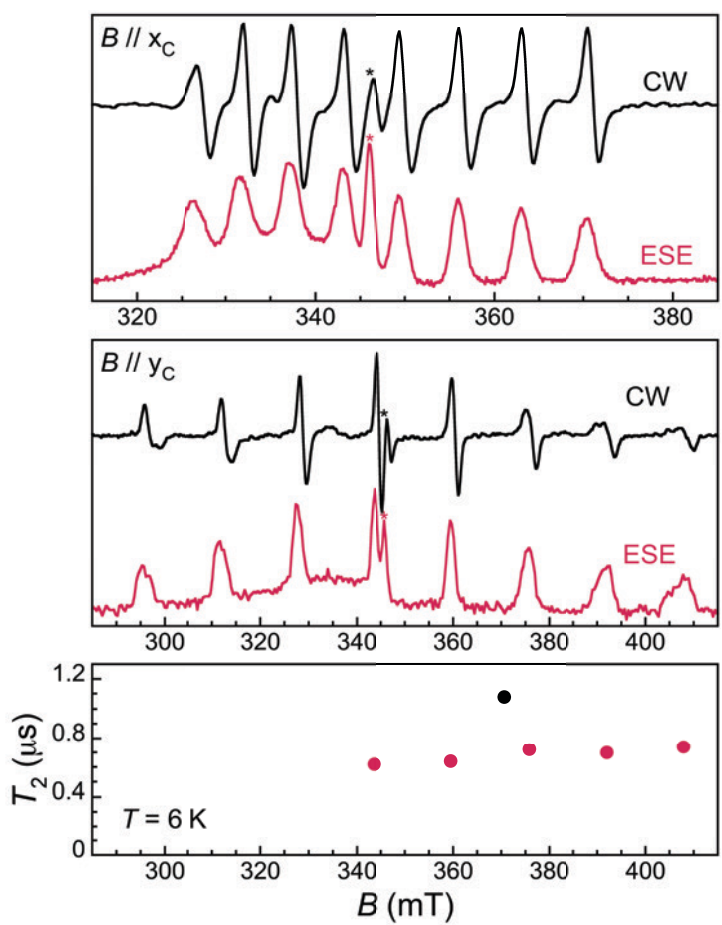

**Figure 4.** *Top*: X-band CW (RT, black lines) and 2p-ESE-detected (6 K, τ = 200 ns, red lines) EPR spectra of **1$^{VO}_{3\%}$** with the field along the x (top) and y (middle) crystal axes. *Bottom*: coherence time $T_2$ for **1$^{VO}_{3\%}$** with the field along the $x_C$ (black symbol) and $y_C$ (red symbols) axes of the crystal. Stars indicate an isotropic signal at g = 2.01 adscribed to a Ti(III) impurity. A broad signal covering the 290-350 mT field range arises from Cu(II) ions on the surface of the cavity.

To evaluate the spin coherence times $T_2$ of the [VOTCPPEt] molecule, we again turn to TD EPR measurements at X-band on both the frozen solution **1$^{VO}_{sol}$** and a diluted single crystal **1$^{VO}_{3\%}$**. The latter were done for two orientations of the crystal, namely along the two perpendicular orientations $x_C$ and $y_C$ for which the splitting of the EPR spectrum is, respectively, minimal and maximal (see Fig. 2). Using a Hahn sequence, consisting of a π/2 pulse and a π pulse separated by a varying interval τ, an intense electron spin echo (ESE) is detected at any magnetic field in the CW-EPR spectrum, indicative of a measurable quantum coherence. The echo-induced EPR spectra are shown in Figs. 2 and 4. They are in excellent agreement with the CW-EPR spectra, being reproduced with the same Hamiltonian parameters given above (see also ESI).

Measurements of the ESE decay as a function of τ were then applied to determine the temperature dependence of the phase memory time $T_2$ in **1$^{VO}_{sol}$**. The decay (Fig. S12) shows a significant modulation which likely arises from the interaction of the vanadyl electronic spin with deuterons from the solvent, in line with recent observations on frozen solutions of Gd complexes.[12b] A low frequency modulation is also present that makes the correct estimation of $T_2$ difficult (see ESI for details). The values of $T_2$ given in Fig. 3, left panel, were derived by fitting a conventional stretched exponential expression, $y(t) = A_0 + A_1\exp\{-(2\tau/T_2)^\beta\}$, to the ESE decays.

The coherence time of [VOTCPPEt] in $\mathbf{1^{VO}_{sol}}$ increases from 3.8(2) ms at 6 K to a plateau at *ca.* 7 ms in the range 25-60 K, and then smoothly decreases down to 1.9(2) ms at 120 K. This decrease is likely due to a parallel decrease of $T_1$, which is only about 1 order of magnitude higher than $T_2$ over the corresponding temperature range. The origin of the increase of $T_2$ observed at low temperature is less clear, and could be an effect of the low frequency modulation, although a similar behavior has been reported for a vanadyl dicathechol complex.[20] In any case, the observed quantum coherence compares well with those found for other $V^{IV}$ molecular spin qubits in frozen solution, such as [V(C$_8$S$_8$)$_3$] (7 ms at 10 K)[22] or [VO(naph-cat$_2$)]$^{2-}$ (6.5 ms at 25 K).[20] The [VOTCPPEt] molecule is therefore a valid spin-qubit candidate, especially considering that the solvent mixture used here is only partly deuterated.

In the solid-state, the ESE decay observed for a single crystal of $\mathbf{1^{VO}_{3\%}}$ at 6 K is also strongly modulated, with an unassigned set of modulation frequencies (Figs. S16 and S17). The quantum coherence of the different electronuclear transitions are very similar, with only a slight increase of $T_2$ with the applied field, from 0.62(2) ms for the central $m_I = -1/2$ to $m_I = 1/2$ transition at 343.6 mT to 0.74(4) ms for the external $m_I = 5/2$ to $m_I = 7/2$ transition at 408.9 mT, all for the magnetic field oriented along $y_c$ (maximum splitting of the EPR spectrum). When the field is applied along $x_c$, *i.e.* for the minimal splitting of the EPR spectrum, the $m_I = 5/2$ to $m_I = 7/2$ transition sees its coherence increase by almost 50% to 1.1(1) ms. While this is significantly lower than in $\mathbf{1^{VO}_{sol}}$, this is likely due to the still relatively high magnetic concentration of $\mathbf{1^{VO}_{3\%}}$. One would therefore expect that $T_2$ approaches that measured in $\mathbf{1^{VO}_{sol}}$ upon further magnetic dilution.

**Broadband on-chip magnetic spectroscopy**

The above characterization has been completed by means of experiments performed on superconducting transmission waveguides coupled to pure crystals of $\mathbf{1^{VO}}$. These experiments provide direct spectroscopic information at any frequency below 14 GHz and as a function of magnetic field and allow varying temperature over a wide range.[12a,23] Results obtained at 4.2 K for different orientations of the magnetic field are shown in Fig. 5 and Fig. S29. The presence of resonant spin transitions manifests in a decrease in the microwave signal transmitted through the device. The normalized transmission (see ESI) is equivalent to the absorption derivative measured in conventional CW EPR experiments. Only, in this case one can sweep both frequency and magnetic field independently of each other, as shown in Fig. 5.

The 2D transmission spectrum provides then a full picture of the Zeeman energy diagram and shows that, for $B > 0.1$ T, the allowed transitions correspond mainly to the reversal of the electronic spin for each nuclear spin projection $m_I$. The hyperfine interaction acts here just as a kind of bias field that shifts resonances associated with different $m_I$ states. In fact, one can directly "read out" the hyperfine level splittings by looking at the frequency shifts between the 8, close to equidistant, resonance lines. This also confirms that the nuclear quadrupolar interaction is close to negligible, in agreement with previous reports for vanadium in diverse configurations.[24]

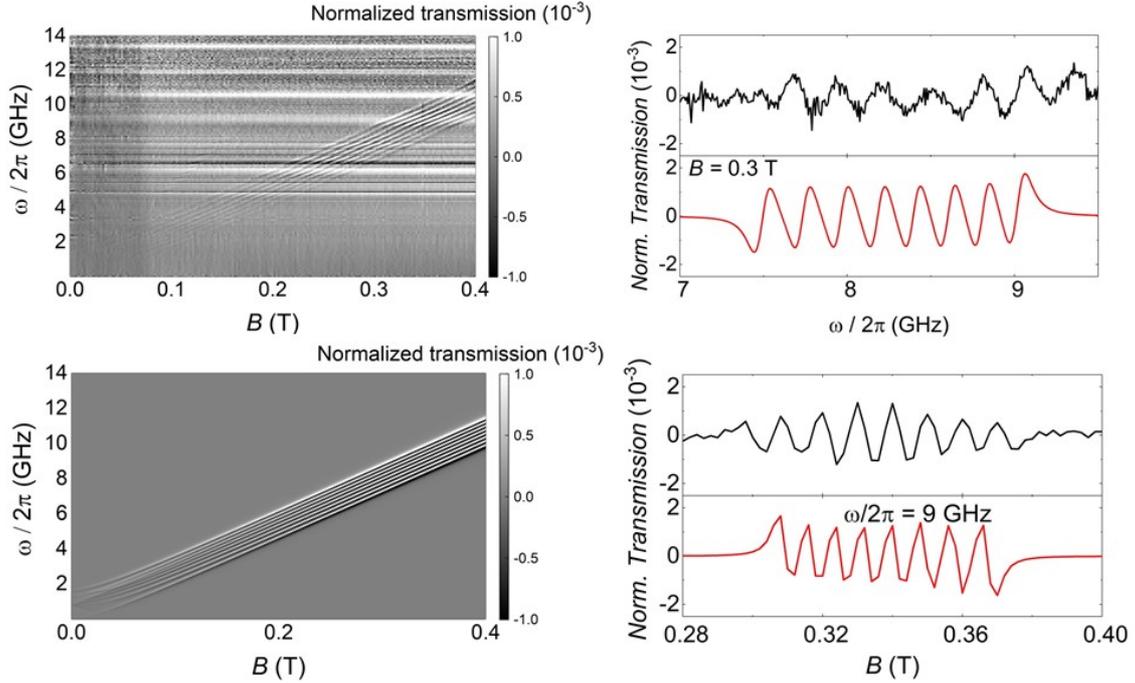

**Figure 5.** *Left*: Grey scale experimental (top) and simulated (bottom) plots of the normalized transmission through a superconducting wave guide coupled to a single crystal of **1$^{VO}$** measured at *T* = 4.2 K as a function of frequency and magnetic field. The magnetic field was applied along the **a** crystallographic axis (X axis of the laboratory frame as defined in Fig. S3). *Right*: Normalized CW spectra obtained from these data at either fixed *B* = 0.3 T as a function of frequency (top) or at fixed $\omega/2\pi$ = 9 GHz as a function of magnetic field. Black and red lines show experimental and simulated data, respectively.

In addition, the effective *g*-factor along each orientation is given by the slope of these lines. Repeating this procedure for the *X*, *Y* and *Z* laboratory axes (see Figs. S28 and S29) and using the crystallographic information shown in Fig. 1, one can refine the parameters of the spin Hamiltonian, which confirm those derived from angle-dependent CW-EPR experiments. Simulations of the normalized transmission done with these parameters, shown in Figs. 5 and S29, agree well with the experimental data.

The results discussed up to now confirm that the [VOTCPPEt] molecules, like other vanadyl derivatives, are quite promising qubit candidates. Within this frame, the existence of multiple nuclear spin levels reduces to a choice between 8 different definitions for the qubit basis states '0' and '1'. However, deviations from this simple picture are expected to occur for weaker magnetic fields, when the Zeeman and hyperfine interactions compete in intensity and the electron and nuclear spins become entangled. Eventually, for *B* = 0 eigenstates of the spin Hamiltonian (1) approach states of the total angular momentum ***F*** = ***S*** + ***I***, with *F* = 4 and 3. The question we address in the following is whether these 16 electronuclear spin states can properly encode the basis states of a *d* = 16 qudit, or of 4-qubit processor. The transmission experiments shown in Fig. 5 allow exploring this low-field low-frequency region, and hint at a deviation of the resonance lines from linearity. However, the signal becomes then also very weak, as all levels are nearly equally populated at 4.2 K. Answering the above question clearly calls for similar experiments in the temperature region below 1 K.

**Very low temperature experiments: [VOTCPPEt] as a *d* = 16 electronuclear spin qudit**

Transmission experiments have been extended to the region of very low temperatures by making use of a $^3$He-$^4$He dilution refrigerator equipped with microwave lines. As it is shown in Fig. S30, decreasing temperature gives rise to a sharp increase in the visibility of the absorption lines. A 2D plot of data measured at the lowest attainable temperature *T* = 0.175 K are shown in Fig. 6. Despite the still sizeable noise, the data reveal some interesting features. First, and as expected, resonance lines no longer follow a linear dependence on *B*. The underlying reason is the occurrence of level crossings (and also anticrossings, which become possible because *F* has integer values). A consequence of this, shown by the two panels on the right hand side of Fig. 6, is that the separations between nearest resonances spread in frequency. The results suggest also that some of the lines split, *i.e.* that additional transitions become allowed in this low-field regime. Again, this is not unexpected since the selection rules that operate a high magnetic fields, when states factorize in electronic and nuclear spin projections, no longer apply when *I* and *S* become entangled. Briefly, these results provide direct evidence for a change in the nature of the energy levels and spin states when the anisotropic hyperfine interactions start competing with the coupling to the external magnetic field. As we discuss below, this provides a crucial ingredient to ensure universal operations between the 16 electronuclear states.

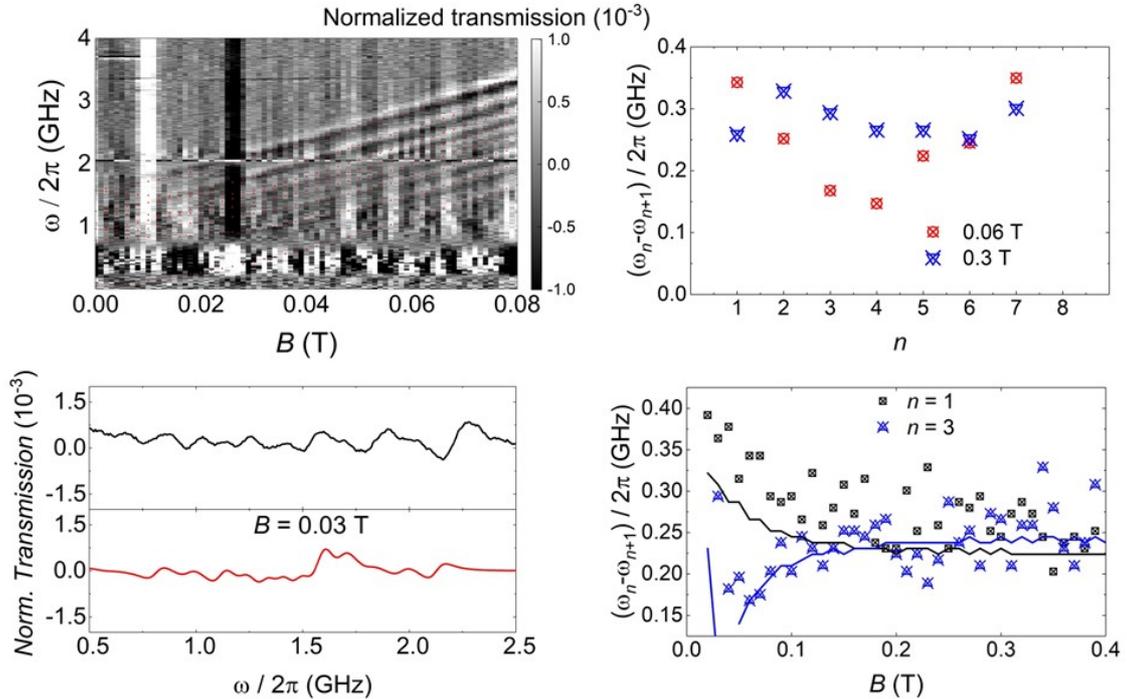

**Figure 6.** *Left, top*: 2D plot of the normalized transmission of a superconducting coplanar transmission waveguide coupled to a single crystal of **1$^{VO}$** measured at *T* = 0.175 K. The magnetic field was parallel to the *Z* laboratory axis. *Left, bottom*: normalized CW spectra obtained from these data at *B* = 0.03 T as a function of frequency. Black and red lines show experimental and simulated data, respectively. *Right*: frequency separations between adjacent resonant lines as a function of either the initial state in each transition (top) or magnetic field (bottom).

In the remaining of this section, we use this information to discuss how to exploit the multiple spin levels of **1$^{VO}$** as additional resources for quantum computation. The conditions that are needed to encode a qudit (or multiple qubits) in an *N*-level system (here $N = (2S+1)\times(2I+1) = 16$) are mainly two. First, one needs to have the possibility of addressing transitions linking different states by properly selecting the frequency of a microwave resonant pulse. This implies that the energy levels must not be equally spaced. And, second, there must be a sufficient number of such transitions that are not forbidden and that have a large enough probability. As we have discussed above, and the level scheme of Fig. 7 shows, degeneracies between inter-level separations are removed, thus the first condition is fulfilled, at sufficiently low magnetic fields. Concerning the second, one must consider the Rabi frequencies $\Omega_R$ of resonant transitions between any pair of spin states. A 2D plot generated for $B_X = 0.02$ T is shown in Fig. 7. It follows from it that not only the "pure" electronic transitions are allowed, but also several others, including those that were previously associated with changes in the nuclear spin state. As *B* decreases, $\Omega_R$ of such "NMR-like" transitions increase at the expense of a similar decrease in $\Omega_R$ values associated with the former (see Fig. 7, right hand panels). The possibility of such an enhanced nuclear spin manipulation has been observed in Yb-trensal molecules by means of pulse-NMR experiments.[9b] For our purposes, the existence of electronuclear entanglement then contributes to generate a much larger set of available operations.

The only remaining, but very important, aspect to consider is whether this set is "sufficiently large". A rigorous criterion is to impose that the system admits universal operations, *i.e.* that one can generate any state within the $d = 16$ dimension Hilbert space via the application of (a sequence of) resonant electromagnetic pulses.[25] This was recently applied to show universality of Gd-based molecular spin qudits.[12] To set a quantitative criterion, we calculate the rates $W_{nm}$ at which any basis state *n* evolves to any other *m*. We require that this evolution is performed by one or several addressable resonant transitions, meaning that their resonant frequencies are nondegenerate with that of any other allowed transition. Since each of them takes a finite evolution time $\Delta t \propto 1/\Omega_R$, the electromagnetic pulses have an intrinsic frequency width proportional to $1/\Delta t$. In order to take this effect into account, we consider only transitions whose frequencies differ by more than $\Omega_R$ and exclude the rest. Next, we impose the condition that all $W_{nm}$ brought about by such resonant transitions (for a microwave field amplitude $b_{mw} = 1$ mT) be larger than a decoherence rate $1/T_2 \approx 0.2$ MHz (or $T_2 = 5$ ms, see Fig. 3). Results obtained for **1$^{VO}$** molecules at two different magnetic fields are given in Fig. 8. The plots show whether any two basis states can be connected by sequences of transitions being both addressable and sufficiently fast.

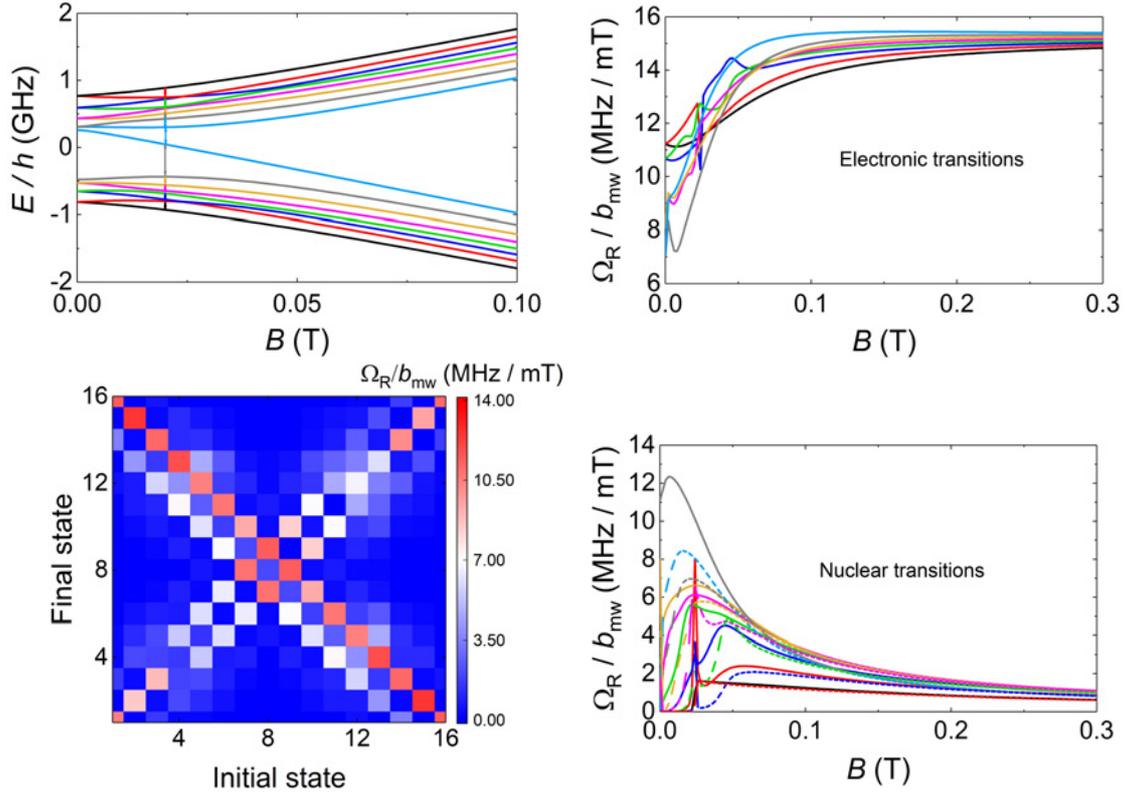

**Figure 7.** *Left, top:* Diagram of electronuclear spin energy levels of [VOTCPPEt] for a magnetic field applied along the X laboratory axis. *Left, bottom*: colour map of Rabi frequencies for resonant transitions, induced by a microwave magnetic field $b_{mw}$ applied along x, linking different electronuclear spin states at $B_X$ = 0.02 T. *Right*: Magnetic field dependence of the Rabi frequencies of "electronic" (top) and "nuclear" (bottom) spin transitions.

The results confirm that **1^VO** affords universal operations at low magnetic fields whereas white spots, signalling the existence of disconnected pairs of states, show up as *B* increases (see also Figs. S34 and S35). At any *B*, the minimum $W_{nm}$ provides a measure of the qudit universality. As can be seen from the bottom of Fig. 8, it goes to 0 as *B* increases, underlying the importance of electronuclear entanglement. Notice, however, that this plot shows the most demanding condition, associated with the slowest operations, typically sequences of NMR transitions. It is likely that the characteristic $T_2$ times of such transitions lie above the average.[9b] Taking this into account, or improving $T_2$, would widen the magnetic field region that warrants universal operation and enable choosing between different situations, with e.g. different resonant frequencies or more or less factorized spin states.

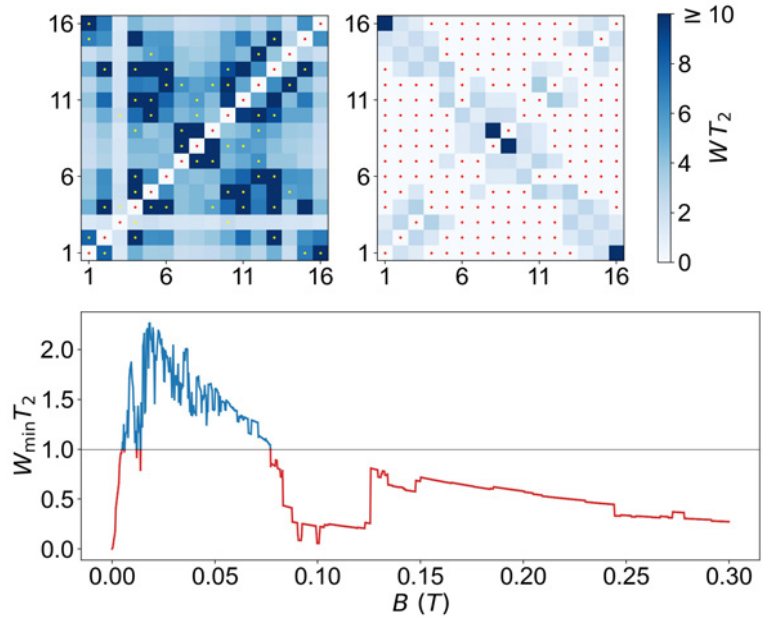

**Figure 8.** Top: Universality plots, showing the rates of operations connecting, via sequences of resonant electromagnetic pulses of amplitude $b_{mw}$ = 1 mT, any pair of electronuclear spin states at $B_X$ = 0.02 T (left) and 0.3 T (right). The coherence time was taken as $T_2$ = 5 ms. White spots, marked also with a red dot, identify two mutually disconnected states and signal that the system cannot implement universal quantum operations. Yellow dots mark the resonant transitions that are used to generate all operations. Bottom: Minimum $W_{nm}T_2$ as a function of $B_X$, showing the transition from universal ($W_{min}$ > 1/$T_2$, blue line) to non-universal ($W_{min}$ < 1/$T_2$, red line) behaviour as the electron and nuclear spin states factorize with increasing magnetic field.

**Conclusions and outlook**

The results discussed in the previous sections show that [VOTCPPEt] molecules provide quite promising realizations of either a 4-qubit processor or a $d$ = 16 qudit. Besides the relatively long spin coherence (of a few ms) and spin relaxation times (up to several s), comparable to that found for some of the best molecular qubits, a key ingredient is the possibility to introduce electronuclear spin entanglement. In [VOTCPPEt], this is done by adequately tuning the relative intensities of hyperfine and Zeeman interactions via the external magnetic field. As a result, resonant transition frequencies become non degenerate and both electronic and nuclear spin transitions can be coherently manipulated at sufficiently high, even comparable, speeds. These conditions ensure that this simple molecular system can perform universal quantum operations. This is a strong statement, meaning that it can provide a suitable implementation for any quantum algorithm.

In Figs. S31-S34, we explore how this result depends on the details of the spin Hamiltonian. It follows that all that is needed to preserve universality is a hyperfine interaction that is not fully uniaxial nor fully isotropic, thus that $A_\perp \neq 0$ and $A_\perp \neq A_\parallel$, while no quadrupolar term is required. The onset of electronuclear entanglement manifests itself (see Fig. 7) in the form of level anticrossings. It is a curious fact that the presence of such "spin-clock transitions" is here exploited as a way to enhance the connectivity between different states, rather than to reduce their sensitivity to noise.[4] This finding widens the "catalogue", and simplifies the search, of molecular systems that can implement non-trivial quantum functionalities, *i.e.* that can contribute to move beyond single qubits. A practical advantage is the possibility of choosing ions, such as the vanadyl center of [VOTCPPEt], with a close to 100% abundance of the relevant (high spin) nuclear isotope, thus avoiding the need, and cost, of isotopical purification.

Embedding nontrivial quantum logic in an *N*-level molecule opens the way to proof-of-concept implementations of some simple algorithms. Examples include Grover's search algorithm, which was recently realized in a 4-level nuclear spin qudit,[10a] and quantum simulations.[10b] But probably the most far-reaching application is the integration of quantum error correction at the molecular level. Four qubits provide a suitable basis for the simplest repetition code, based on Shor's algorithm,[9a] able to correct for single qubit or single phase flips. However, a better alternative seems to be the use of codes adapted to the level structure of qudits, which not even require that *N* be a power of 2. A particularly efficient code has been recently developed for spin qudits coupled to ancillary $S = 1/2$ electron spins.[9b,9c] Increasing the qudit dimension helps improving the state fidelity for any fixed decoherence rate. Each [VOTCPPEt] molecule, with its $I = 7/2$ vanadium qudit coupled to the $S = ½$ vanadyl electronic spin, provides a promising platform to implement such a code.

Using molecules as error-protected logical qubits to build increasingly complex quantum devices underlines one of the competitive advantages that chemical design can bring to the development of scalable quantum technologies. Yet, this development still faces important challenges, mainly connected with the control of individual molecular spins. Still, [VOTCPPEt] and other similar systems might allow implementing proof-of-concept realizations of these codes by much more easily affordable experiments on crystals. The experiments reported here show that nearly all of the required ingredients can be met in practice. First, it is possible to attain sufficiently long spin coherence times in diluted single crystals (see Fig. 4), without losing the spatial organization of the molecules, *i.e.* while keeping the possibility of selectively addressing individual transitions. This possibility has been recently used to implement single qubit operations on a related system by means of on-chip resonators.[27] In order to exploit the multiple spin states, the coupling to open transmission lines allows generating broad-band control electromagnetic pulses, at the cost of reducing the electromagnetic field amplitude. In the case of [VOTCPPEt], relevant frequencies range between a few hundred MHz up to 2 GHz at low fields. For this region of frequencies, microwave pulse amplitudes of about 1 mT can be easily generated, and this value can be improved by adequate circuit design.[26] This gives rise to Rabi frequencies for individual operations that are in the range of a few (tens of) MHz (see Fig. 7), thus faster than decoherence rates. Finally, the use of qudit codes simplifies also reading-out the result of the quantum operations, as it can be obtained by measuring the resonance of the electronic spin. For this, a transmission line can be combined with on-chip LC resonators with frequencies in the range of 1-2 GHz. In conclusion, these results can contribute to bring molecular spins closer to practical quantum implementations.

**Conflicts of interest**

There are no conflicts to declare.

**Acknowledgements**

This work was supported by funds from the EU (COST Action 15128 MOLSPIN, QUANTERA project SUMO, FET-OPEN grant 862893 FATMOLS), the Spanish MICINN (grants CTQ2015-68370-P, CTQ2015-64486-R, RTI2018-096075-B-C21, PCI2018-093116, PGC2018-098630-B-I00, MAT2017-86826-R) and the Gobierno de Aragón (grants E09-17R-Q-MAD, E31_17R PLATON).

**Notes and references**
1 a) T. D. Ladd, F. Jelezko, R. Laflamme, Y. Nakamura, C. Monroe and J. L. O'Brien, *Nature*, 2010, **464**, 45; b) A. Gaita-Ariño, F. Luis, S. Hill and E. Coronado, *Nat. Chem.*, 2019, **11**, 301; c) M. Atzori and R. Sessoli, *J. Am. Chem. Soc.*, 2019, **141**, 11339–11352; d) G. Aromí and O. Roubeau, *Handbook of Physics and Chemistry of Rare-Earths*, 2019, **59**, 1

2 avoiding protons and rotating methyl groups in a [$Cr_7Ni$] ring, see C. J. Wedge, G. A. Timco, E. T. Spielberg, R. E. George, F. Tuna, S. Rigby, E. J. L. McInnes, R. E. P. Winpenny, S. J. Blundell and A. Ardavan, *Phys. Rev. Lett.*, 2012, **108**, 107204

3 avoiding nuclear spins in general, see K. Bader, D. Dengler, S. Lenz, B. Endeward, S-D. Jiang, P. Neugebauer and J. van Slageren, *Nat. Commun.*, 2015, **5**, 5304

4 using clock transitions see a) M. Shiddiq, D. Komijani, Y. Duan, A. Gaita-Ariño, E. Coronado and S. Hill, *Nature*, 2016, **531**, 348; b) J. M. Zadrozny, A. T. Gallagher, T. D. Harris and D. E. Freedman, *J. Am. Chem. Soc.*, 2017, **139**, 7089

5 M. Jenkins, T. Hümmer, M. J. Martínez-Pérez, J. J. García-Ripoll, D. Zueco and F. Luis, *New J. Phys.*, 2013, **15**, 095007

6 a) S. Thiele, F. Balestro, R. Ballou, S. Klyatskaya, M. Ruben, W. Wernsdorfer, *Science*, 2014, **344**, 1135-1138; b) J. Liu, J. Mrozek, W. K. Myers, G. A. Timco, R. E. P. Winpenny, B. Kintzel, W. Plass and A. Ardavan, *Phys. Rev. Lett.*, 2019, **122**, 037202

7 a) M. S. Fataftah, S. L. Bayliss, D. W. Laorenza, X. Wang, B. T. Phelan, C. B. Wilson, P. J. Mintun, B. D. Kovos, M. R. Wasielewski, S. Han, M. S. Sherwin, D. D. Awschalom and D. E. Freedman, *J. Am. Chem. Soc.*, 2020, **142**, 20400; b) S. L. Bayliss, D. W. Laorenza, P. J. Mintun, B. D. Kovos, D. E. Freedman and D. D. Awschalom, *Science*, 2020, **370**, 1309

8 a) G. Aromí, D. Aguilà, P. Gamez, F. Luis and O. Roubeau, *Chem. Soc. Rev.*, 2012, **41**, 537; b) E. Moreno-Pineda, C. Godfrin, F. Balestro, W. Wernsdorfer and M. Ruben, *Chem. Soc. Rev.*, 2018, **47**, 501-513

9 a) E. Macaluso, M. Rubín, D. Aguilà, A. Chiesa, L. A. Barrios, J. I. Martínez, P. J. Alonso, O. Roubeau, F. Luis, G. Aromí and S. Carretta, *Chem. Sci.*, 2020, **11**, 10337-10343; b) R. Hussain, G. Allodi, A. Chiesa, E. Garlatti, D. Mitcov, A. Konstantatos, K. S. Pedersen, R. De Renzi, S. Piligkos and S. Carretta, *J. Am. Chem. Soc.*, 2018, **140**, 9814–9818; c) A. Chiesa, E. Macaluso, F. Petiziol, S. Wimberger, P. Santini and S. Carretta, *J. Phys. Chem. Lett.*, 2020, **11**, 8610–8615

10 a) C. Godfrin, A. Ferhat, R. Ballou, S. Klyatskaya, M. Ruben, W. Wernsdorfer and F. Balestro, *Phys. Rev. Lett.*, 2017, **119**, 187702; b) A. Chiesa, G. F. S. Whitehead, S. Carretta, L. Carthy, G. A. Timco, S. J. Teat, G. Amoretti, E. Pavarini, R. E. P. Winpenny and P. Santini, *Sci. Reports*, 2014, **4**, 7423; c) M. Atzori, A. Chiesa, E. Morra M. Chiesa, L. Sorace S. Carretta and R. Sessoli, *Chem. Sci.*, 2018, **9**, 6183-6192 11 a) F. Luis, A. Repollés, M. J. Martínez-Pérez, D. Aguilà, O. Roubeau, D. Zueco, P. J. Alonso, M. Evangelisti, A. Camón, J. Sesé, L. A. Barrios and G. Aromí, *Phys. Rev. Lett.*, 2011, **107**, 117203; b) D. Aguilà, L. A. Barrios, V. Velasco, O. Roubeau, A. Repollés, P. J. Alonso, J. Sesé, S. J. Teat, F. Luis and G. Aromí, *J. Am. Chem. Soc.*, 2014, **136**, 14215; c) A. Ardavan, A. M. Bowen, A. Fernández, A. J. Fielding, D. Kaminski, F. Moro, C. A. Muryn, M. D. Wise, A. Ruggi, E. J. L. McInnes, K. Severin, G. A. Timco, C. R. Timmel, F. Tuna, G. F. S. Whitehead and R. E. P. Winpenny, *npj Quantum Inf.*, 2015, **1**, 15012; d) J. Ferrando-Soria, E. M. Pineda, A. Chiesa, A. Fernández, S. A. Magee, S. Carretta, P. Santini, I. J. Vitorica-Yrezabal, F. Tuna, G. A. Timco, E. J. L. McInnes and R. E. P. Winpenny, *Nat. Commun.*, 2016, **7**, 11377; e) J. Salinas Uber, M. Estrader, J. Garcia, P. Lloyd-Williams, A. Sadurní, D. Dengler, J. van Slagaren, N. F. Chilton, O. Roubeau, S. J. Teat, J. Ribas-Ariño and G. Aromí, *Chem. Eur. J.*, 2017, **23**, 13648; f) J. Ferrando-Soria, S. A. Magee, A. Chiesa, S. Carretta, P. Santini, I. J. Vitorica-Yrezabal, F. Tuna, G. F. S. Whitehead, S. Sproules, K. M. Lancaster, A-L. Barra, G. A. Timco, E. J. L. McInnes and R. E. P. Winpenny, *Chem*, 2016, **1**, 727
12 a) M. D. Jenkins, Y. Duan, B. Diosdado, J. J. García-Ripoll, A. Gaita-Ariño, C. Giménez-Saiz, P. J. Alonso, E. Coronado and F. Luis, *Phys. Rev. B.*, 2017, **95**, 064423; b) F. Luis, P. J. Alonso, O. Roubeau, V. Velasco, D. Zueco, D. Aguilà, J. I. Martínez, L. A. Barrios and G. Aromí, *Commun. Chem.*, 2020, **3**, 176
13 E. Moreno-Pineda, M. Damjanović, O. Fuhr, W. Wernsdorfer and M. Ruben, *Angew. Chem.*, 2017, **56**, 9915-9919
14 M. D. Jenkins, D. Zueco, O. Roubeau, G. Aromí, J. Majer, and F. Luis, *Dalton Trans.*, 2016, **45**, 16682
15 M. Atzori, L. Tesi, E. Morra, M. Chiesa, L. Sorace and R. Sessoli, *J. Am. Chem. Soc.*, 2016, **138**, 2154
16 J. M. Zadrozny, J. Niklas, O. G. Poluektov and D. E. Freedman, *ACS Central Sci.*, 2015, **1**, 488
17 A. Urtizberea, E. Natividad, P. J. Alonso, L. Pérez-Martínez, M. A. Andrés, I. Gascón, I. Gimeno, F. Luis and O. Roubeau, *Mater. Horiz.*, 2020, **7**, 885
18 D. C. Johnston, *Phys. Rev. B*, 2006, **74**, 184430
19 a) F. Luis, F. L. Mettes, J. Tejada, D. Gatteschi and L. J. de Jongh, *Phys. Rev. Lett.* 2000, **85**, 4377;  b) M. Evangelisti, F. Luis, L. J. de Jongh and M. Affronte, *J. Mater. Chem.*, 2006, **16**, 2534–2549
20 M. Atzori, S. Benci, E. Morra, L. Tesi, M. Chiesa, R. Torre, L. Sorace and R. Sessoli, *Inorg. Chem.*, 2018, **57**, 731
21 K. Bader, M. Winkler, and J. Van Slageren, *Chem. Commun.*, 2016, **52**, 3623
22 C-J. Yu, M. J. Graham, J. M. Zadrozny, J. Niklas, M. D. Krzyaniak, M. R. Wasielewski, O. G. Poluektov and D. E. Freedman, *J. Am. Chem. Soc.*, 2016, **138**, 14678
23 C. Clauss, D. Bothner, D. Koelle, R. Kleiner, L. Bogani, M. Scheffler and M. Dressel, *Appl. Phys. Lett.*, 2013, **102**, 162601
24 a) K. Paulsen and D. Rehder, *Z. Naturforsch.*, 1982, **37a**, 139; b) T. S. Smith II, R. LoBrutto and V. L. Pecoraro, *Coord. Chem. Rev.*, 2002, 228, 1; c) C. P. Aznar, Y. Deligiannakis, E. J. Tolis, T. Kabanos, M. Brynda and R. D. Britt, *J. Phys. Chem. A*, 2004, **108**, 4310
25 C. V. Kraus, M. Wolf, J. I. Cirac, *Phys. Rev. A*, 2007, **75**, 9

26 a) A. Bienfait, J. J. Pla, Y. Kubo, M. Stern, X. Zhou, C. C. Lo, C. D. Weis, T. Schenkel, M. L. W. Thewalt, D. Vion, D. Esteve, B. Julsgaard, K. Mølmer, J. J. L. Morton and P. Bertet, *Nature Nanotechnology*, 2016, **11**, 253; b) B. Sarabi, P. Huang and N. M. Zimmerman, *Phys. Rev. Appl.*, 2019, **11**, 014001; c) I. Gimeno, W. Kersten, M. C. Pallarés, P. Hermosilla, M. J. Martínez-Pérez, M. D. Jenkins, A. Angerer, C. Sánchez-Azqueta, D. Zueco, J. Majer, A. Lostao and F. Luis, *ACS Nano*, 2020, **14**, 8707

27 C. Bonizzoni, A. Ghirri, F. Santanni, M. Atzori, L. Sorace, R. Sessoli and Marco Affronte, *npj Quantum Information*, 2020, **6**, 68

Supporting Information for the manuscript:

**Broad-band spectroscopy of a vanadyl porphyrin: a model electronuclear spin qudit**

Ignacio Gimeno, Ainhoa Urtizberea, Juan Román-Roche, David Zueco, Agustín Camón, Pablo J. Alonso, Olivier Roubeau,* and Fernando Luis*

## Table of contents

**Synthesis**

Commercial 5,10,15,20-tetrakis(4-carboxyphenyl)porphine (H$_6$TCPP, >97%) was purchased from TCI. VO(SO$_4$)·H$_2$O (>99.0%), Ti(IV)oxysulfate (>29% Ti as TiO$_2$ basis), were purchased from Aldrich and used without further purification.

**Synthesis of [VO(TCPPEt)] (1$^{VO}$).** This was done as previously reported,[1] under conditions that were used before for the Cu analogue.[2] H$_6$TCPP (81 mg, 0.10 mmol), VO(SO$_4$)·H$_2$O (17 mg, 0.094 mmol) and 10 mL ethanol were gently mixed in a 23 mL Teflon-lined PARR acid digestion bomb, and the operation repeated 5 times. The five bombs were placed in an oven at 60ºC, warmed to 180ºC and kept at this temperature for 24 h. After cooling to RT, large shiny violet crystals were recovered by filtration, washed with little ethanol and dried in air. The total yield of crystals of 1$^{VO}$ was 422 mg (77% based on V, 0.436 mmol). MALDI-TOF-MS (matrix: CHCA): m/z = 967.6. EA calc. (found) for C$_{56}$H$_{44}$N$_4$O$_9$V: C, 69.49 (69.2); H, 4.58 (4.5); N, 5.79 (5.6) wt%.

**Synthesis of [(TiO)$_{(1-x)}$(H$_{2x}$TCPPEt)] (1$^{TiO}$).** H$_6$TCPP (84 mg, 0.104 mmol), TiO(SO$_4$)·xH$_2$O (24 mg, 0.139 mmol considering x = 1) and 10 mL ethanol were gently mixed in a 23 mL Teflon-lined PARR acid digestion bomb, and the bomb was placed in an oven at 60ºC, warmed to 180ºC and kept at this temperature for 24 h. After cooling to RT, large shiny violet crystals were recovered by filtration, washed with little ethanol and dried in air. The total yield of crystals of 1$^{TiO}$ was 68 mg. Repeated reactions systematically gave crystals with only a partial metallation. The Ti content determined through ICP-MS was reproducibly corresponding to x ranging from 0.4 to 0.6, in agreement with repeated single-crystal structure refinements. The reported structure for 1$^{TiO}$ refines to x = 0.5. We did not find any crystal of the free-base H$_2$TCPPEt over 20+ single-crystals tested.

The same reaction was repeated with increasing excess of TiO(SO$_4$)·xH$_2$O up to *ca.* 5-fold excess (103 mg, 0.506 mmol considering x = 1), resulting in a mixture of large violet crystals and an increasing amount of white solid, found to be unreacted TiO(SO$_4$)·xH$_2$O based on IR spectra. The violet crystals were found to be either crystals of 1$^{TiO}$, still with x of the order of 0.5 or crystals of the free-base H$_2$TCPPEt, from which the reported structure was determined.

**Synthesis of [(VO)$_{0.03}$(TiO)$_{0.53}$(H$_2$)$_{0.44}$TCPPEt)] (1$^{VO}$$_{3\%}$).** The synthesis was done as for 1$^{TiO}$ albeit using H$_6$TCPP (83 mg, 0.103 mmol), VO(SO$_4$)·H$_2$O (0.6 mg, 0.003 mmol) and TiO(SO$_4$)·xH$_2$O (20.3 mg, 0.118 mmol considering x = 1). Large shiny violet crystals were recovered by filtration, washed with little ethanol and dried in air. The total yield of crystals of 1$^{VO}$$_{3\%}$ was 75 mg. Single-crystal X-ray diffraction repeatedly gave a unit-cell very close to that found for 1$^{TiO}$ (see Table S1). The average metal composition was determined by ICP-MS. As expected from the outcome of the synthesis of 1$^{TiO}$, metallation by the titanyl ion is not complete and the crystals are a mixture of vanadyl, titanyl and unmetallated porphyrins. Each crystal likely has a slightly different composition.

**Synthesis of H$_2$TCPPEt.** H$_6$TCPP (84 mg, 0.104 mmol) and 10 mL ethanol were gently mixed in a 23 mL Teflon-lined PARR acid digestion bomb, and the bomb was placed in an oven at 60ºC, warmed to 180ºC and kept at this temperature for 24 h. After cooling to RT, a violet polycrystalline powder was recovered by filtration, washed with little ethanol and dried in air. The total yield of H$_2$TCPPEt was 58 mg. No large single-crystal were obtained, and attempts to obtain large single-crystals of [(VO)$_x$(H$_2$)$_{2-2x}$TCPPEt] also only yielded polycrystalline powders.

---

[1] A. Urtizberea, E. Natividad, P. J. Alonso, L. Pérez, M. A. Andrés, I. Gascón, I. Gimeno, F. Luis, O. Roubeau, *Mater. Horiz.*, **2020**, 7, 885.

[2] W. Chen, and S. Fukuzumi, *Eur. J. Inorg. Chem.*, **2009**, 5494.

**Single-Crystal X-Ray Crystallography (SCXRD)**
Data for H$_2$TCPPEt were acquired at 100 K on a 0.20x0.17x0.16 mm$^3$ violet block on a Bruker APEX II QUAZAR diffractometer equipped with a microfocus multilayer monochromator with MoK$\alpha$ radiation ($\lambda$ = 0.71073 Å). Data reduction and absorption corrections were performed with respectively SAINT and SADABS.[3] Data for **1$^{VO}$** and **1$^{TiO}$** were obtained respectively on a 0.88x0.47x0.19 mm$^3$ violet block at 150 K and on a 0.22x0.06x0.03 mm$^3$ violet lath at 295 K on an Oxford Diffraction Excalibur Sapphire3 diffractometer with enhanced MoK$\alpha$ radiation ($\lambda$ = 0.71073 Å), at the X-ray diffraction and Fluorescence Analysis Service of the University of Zaragoza. Cell refinement, data reduction, and absorption corrections were performed with the CrysAlisPro suite.[4] All structures were solved by intrinsic phasing with SHELXT[5] and refined by full-matrix least-squares on $F^2$ with SHELXL-2014.[6] All details can be found in CCDC 2058759-2058760-2058761 (H$_2$TCPPEt-**1$^{VO}$**-**1$^{TiO}$**) that contain the supplementary crystallographic data for this paper. These data can be obtained free of charge from The Cambridge Crystallographic Data Center via https://summary.ccdc.cam.ac.uk/structure-summary-form. Crystallographic and refinement parameters are summarized in Table S1. Selected bond lengths and angles and intermolecular distances are given in Tables S2 and S3.

**Electron Paramagnetic Resonance (EPR)** experiments, both continuous wave (CW) and pulsed time domain (TD), were performed with a Bruker Biospin ELEXSYS E-580 spectrometer operating in the X-band, using a gas-flow Helium cryostat for low-temperature experiments. Rotational CW studies were on single-crystals of both **1$^{VO}$** and **1$^{VO}_{3\%}$** mounted on 2x2x2 mm$^3$ cube-shaped holder (see Figure S3). TD experiments were done on a frozen solution **1$^{VO}_{sol}$** (0.46 mmol/L **1$^{VO}$** in 1:1 mixture of toluene and CDCl$_3$) and on a single-crystal of **1$^{VO}_{3\%}$**. 2p and Inversion Recovery ESE-detected experiments were performed. The simulated spectra were obtained with the EasySpin program.[7]

**Magnetothermal characterization**
Magnetization and susceptibility data of polycrystalline **1$^{VO}$** were measured, between 1.8 K and 300 K and for dc magnetic fields up to 5 T, with a commercial magnetometer equipped with a SQUID sensor hosted by the Physical Measurements Unit of the Servicio General de Apoyo a la Investigación-SAI, Universidad de Zaragoza. This system was also used to perform ac susceptibility measurements. For this, a sinusoidal ac magnetic field with amplitude $b_0$ = 4 Oe, oscillating at a frequency $\omega/2\pi$ between 1 Hz and 1.1 kHz, was applied parallel to the static magnetic field and the linear in-phase $\chi'$ and out-of-phase $\chi''$ susceptibility components were recorded. Additional ac susceptibility measurements were performed in the range 10 ≤ $\nu$ ≤ 10000 Hz using the ACMS option of a commercial physical property measurement system. The diamagnetic contributions to the magnetization and susceptibility were corrected using Pascal's constant tables.

Additional magnetization measurements were performed, between 0.35 K and 5 K, on a single crystal of **1$^{VO}$** using a home-made micro-Hall magnetometer. The sample was placed on the edge of one of the three Hall crosses. A magnetic field $B$ < 2 T was applied along the plane of the sensor to minimize its intrinsic bare signal. This signal was calibrated and then subtracted from the results.

---

[3] G. M. Sheldrick, *SAINT and SADABS*, Bruker AXS Inc.: Madison, Wisconsin, USA, 2012.
[4] *CrysAlis PRO*, Agilent Technologies Ltd, Yarnton, Oxfordshire, England.
[5] G. M. Sheldrick, *Acta Cryst. A* **2015**, *71*, 3-8.
[6] G. M. Sheldrick, *Acta Cryst. C* **2015**, *71*, 3-8.
[7] S. Stoll, A. Schweiger, *J. Magn. Reson.*, **2006**, *178*, 42.

The specific heat of $\mathbf{1^{VO}}$ was measured, between $T$ = 0.35 K and 100 K, with a commercial physical property measurement system (PPMS, of the Servicio de Apoyo a la Investigación-SAI, Universidad de Zaragoza). A single crystal was placed onto a thin layer of apiezon N grease that fixes the sample and improves the thermal contact with the calorimeter. The PPMS makes use of the relaxation method, which measures the temperature evolution that follows after the application and subsequent removal of a heat power pulse. Experiments were performed with different durations of the pulse, to explore the dynamics in the thermalization of vanadyl spins. The raw data were corrected from the known contributions arising from the empty calorimeter and the grease.

**On-chip magnetic spectroscopy**
The superconducting coplanar waveguides employed in this work consist of a 800 μm wide central transmission line separated from two ground planes by 400 μm wide gaps. They are fabricated by optical lithography on 150 nm thick Nb films ($T_C$ = 9.2 K) deposited by sputtering onto single crystalline sapphire wafers. The size of the central line and its meander shape were designed in order to best match the dimensions of the large $\mathbf{1^{VO}}$ single crystals used in the experiments (an optical microscopy image of a chip can be seen in Figure S26). Microwave transmission experiments were performed on a Helium cryostat ($T$ = 4.2 K) and a $^3$He-$^4$He dilution refrigerator (0.15 K < $T$ < 6 K), both of them compatible with the used 9 T × 1 T × 1 T superconducting vector magnet. The chips input and output lines were connected to a vector network analyser that measures the transmission, $S_{12}$ and $S_{21}$, and reflection, $S_{11}$ and $S_{22}$, coefficients.

The crystals were positioned onto the transmission line with Paraton N grease. The long axis of the crystals, that coincides with the crystallographic 101 axis, was oriented nearly parallel to the laboratory Z and to the axis of the device, which corresponds also to the orientation of the microwave magnetic field seen by the majority of molecules in the crystal (Figures 1 main text and S26). In order to compensate for the decay of the waveguide transmission with increasing frequency and to enhance the contrast of those effects associated with its coupling to the spins, $S_{21}$ and $S_{12}$ were normalized by a method similar to that proposed in reference 8. The normalized transmission $t$ at magnetic field $B_1$ and frequency $\omega_1$ is given by

$$t(B_1, \omega_1) = \frac{S_{21}(B_1, \omega_1) - S_{21}(B_2, \omega_1)}{S_{21}^{(0)}(\omega_1)}$$

where $B_2 > B_1$ and $S_{21}^{(0)}$ is the transmission of the 'empty' transmission line. In practice, $S_{21}^{(0)}$ is measured at a magnetic field for which all spin excitations are outside the accessible frequency region. For $B_2$ very close to $B_1$ (*i.e.* closer than the magnetic field width of a given absorption line), $t$ approximately corresponds to the derivative of the normalized transmission, similar to the signal detected by conventional Electron Paramagnetic Resonance (EPR) systems. The actual transmission can also be obtained by choosing $B_2$ sufficiently far from $B_1$, but at the cost of deteriorating the signal-to noise ratio.

---

[8] I. Gimeno, D. Zueco, Y. Duan, C. Sánchez-Azqueta, T. Astner, A. Gaita-Ariño, S. Hill, J. Majer, E. Coronado and F. Luis, arXiv:1911.07541

**Table S1.** Crystallographic and refinement parameters for the structures of **1**[VO], **1**[TiO] and H$_2$TCPPEt. A representative example of the unit-cells obtained for **1**[VO]$_{3\%}$ is given for comparison.

| Compound | **1**[VO] | **1**[TiO] | **1**[VO]$_{3\%}$ | H$_2$TCPPEt |
|---|---|---|---|---|
| Formula | C$_{56}$H$_{44}$N$_4$O$_9$V | C$_{56}$H$_{44}$N$_4$O$_{8.5}$Ti$_{0.5}$ | | C$_{56}$H$_{46}$N$_4$O$_8$ |
| FW (g mol$^{-1}$) | 967.89 | 933.91 | | 902.97 |
| Wavelength (Å) | 0.71073 | 0.71073 | 0.71073 | 0.71073 |
| $T$ (K) | 150(2) | 295(2) | 295(2) | 100(2) |
| Crystal system | monoclinic | monoclinic | monoclinic | monoclinic |
| Space group | $P2_1/n$ | $P2_1/n$ | | $P2_1/n$ |
| $a$ (Å) | 9.3937(6) | 9.4036(14) | 9.393(3) | 9.0183(15) |
| $b$ (Å) | 10.8559(6) | 10.8957(13) | 10.878(3) | 10.8486(18) |
| $c$ (Å) | 22.317(2) | 22.319(3) | 22.319(6) | 22.331(3) |
| $\beta$ (°) | 93.508(8) | 90.524(13) | 90.73(3) | 92.992(9) |
| $V$ (Å$^3$) | 2271.6(3) | 2286.7(5) | 2280(2) | 2181.8(6) |
| $Z$ | 2 | 2 | | 2 |
| $\rho_{calcd}$ (g cm$^{-3}$) | 1.415 | 1.356 | | 1.374 |
| $\mu$ (mm$^{-1}$) | 0.285 | 0.171 | | 0.093 |
| Reflections | 3997 | 1359 | | 3693 |
| $R_{int}$ | 0.0434 | 0.0664 | | 0.0291 |
| Restraints | 156 | 7 | | 17 |
| Parameters | 385 | 324 | | 326 |
| $S$ | 1.029 | 1.016 | | 1.132 |
| $R_1$ [$I>2\sigma(I)$] | 0.0895 | 0.0672 | | 0.0601 |
| $wR_2$ [$I>2\sigma(I)$] | 0.2274 | 0.1724 | | 0.1536 |
| $R_1$ [all data] | 0.1210 | 0.1054 | | 0.0660 |
| $wR_2$ [all data] | 0.2496 | 0.2071 | | 0.1573 |
| Largest peak / hole ($e$ Å$^3$) | 0.359 / −1.111 | 0.152 / −0.187 | | 0.842 / −0.389 |

**Table S2.** Details of intermolecular C–H···$\pi$ and $\pi$···$\pi$ interactions in the structure of **1**[VO] (see Figs. S1 and S2).

| pyrrole centroid | H–C | distance (Å) |
|---|---|---|
| $C_g$ (N2C7C8C9C10) | H28A–C28 | 2.755 |
| $C_g$ (N1C2C3C4C5) | H25A–C25 | 2.788 |

| pyrrole centroid | ester pivot carbon | distance (Å) |
|---|---|---|
| $C_g$ (N2C7C8C9C10) | C17 | 3.449 |

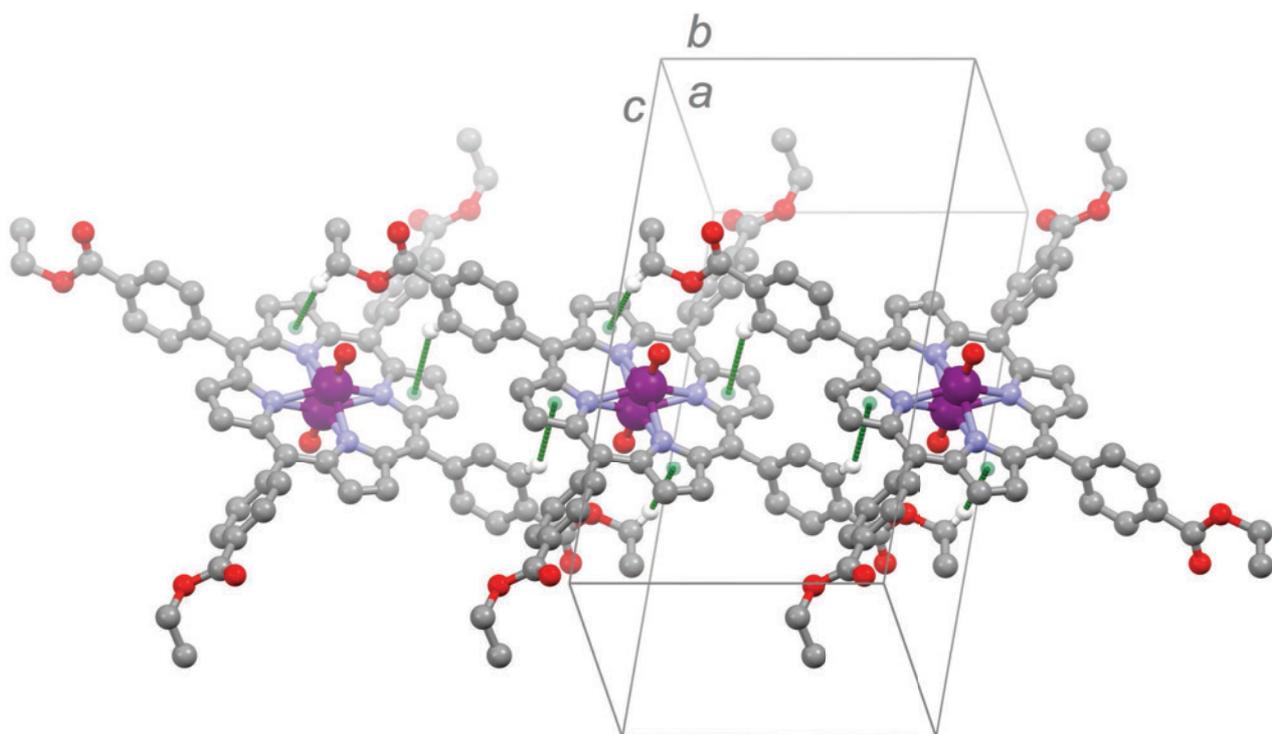

**Figure S1.** Portion of the packing in the structure of **1^VO** highlighting the intermolecular C-H···π interactions (green sticks) giving rise to supramolecular chains of [VOTCPPEt] molecules with identical orientation. Colour code: plum, V; red, O; light blue, N; grey, C, white, H, green, pyrrole rings centroids. Only hydrogens involved in the interactions are shown for clarity. See Table S2 for details.

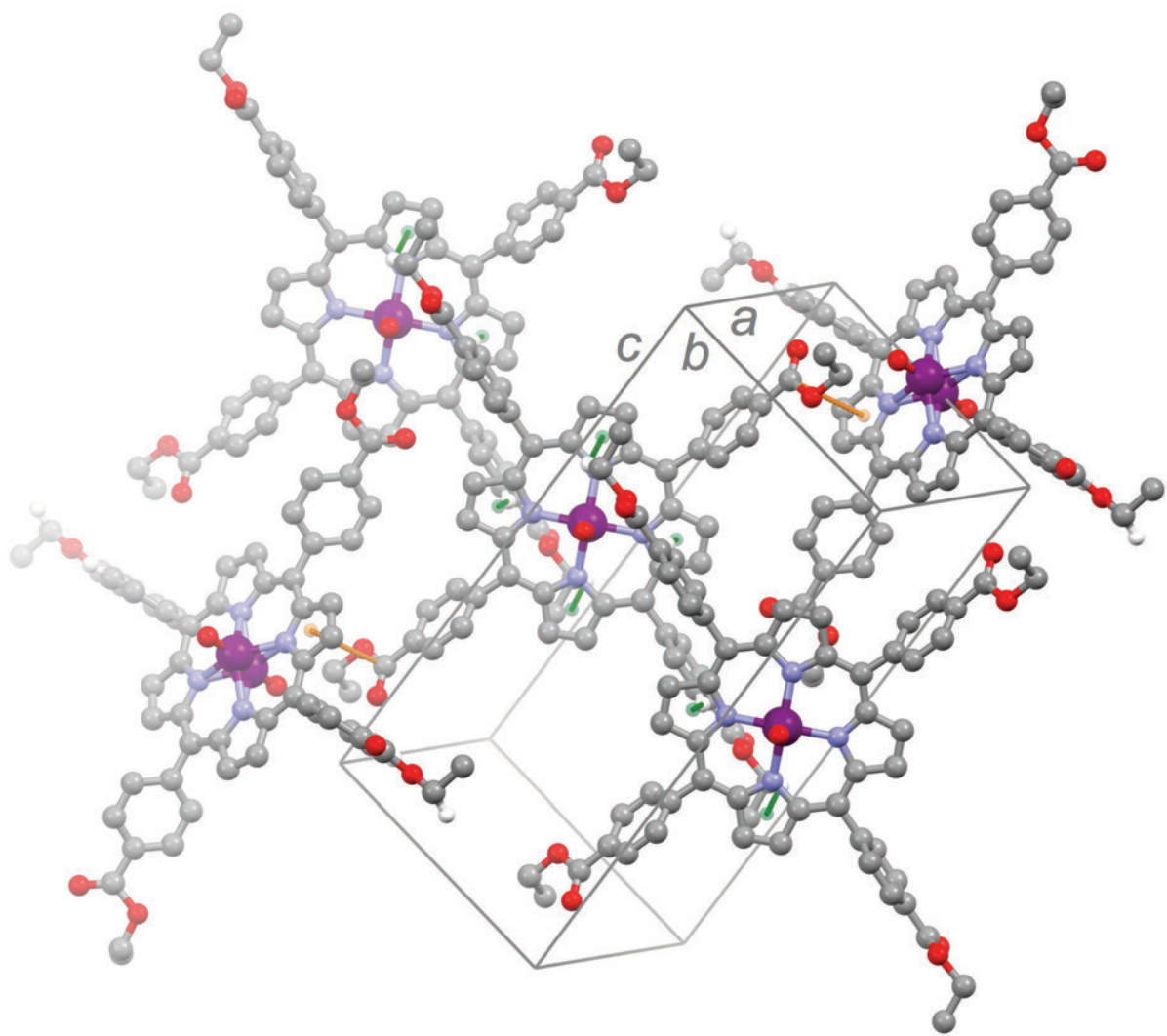

**Figure S2.** Portion of the packing in the structure of **1**[VO] highlighting the intermolecular π⋯π interactions between pyrrole rings and adjacent COO moieties (orange sticks) between [VOTCPP] molecules with different orientations. These connect the supramolecular chains (shown here is the same as in Fig. S1) in planes. Colour code: plum, V; red, O; light blue, N; grey, C, white, H, green and orange, pyrrole rings centroids. Only hydrogens involved in the C-H⋯π interactions (green sticks) are shown for clarity. See Table S2 for details.

**Details and analysis of single-crystal EPR measurements.**

Crystals were mounted on a "perpex" cube-shaped holder (2x2x2 mm³), using GE varnish to avoid any uncontrolled modification of the crystal orientation. Figure S3a shows the crystal of **1$^{VO}_{3\%}$** on the holder after the measurements, and the definition of the laboratory axes X, Y and Z used in experiments. The Z axis coincides with the long growth direction of the crystal, the monoclinic *b* axis, while the XY plane is parallel to the glide plane. The dc magnetic field B and the microwave magnetic field $b_{mw}$ generated by the cavity were perpendicular to the rotation axis and orthogonal to each other.

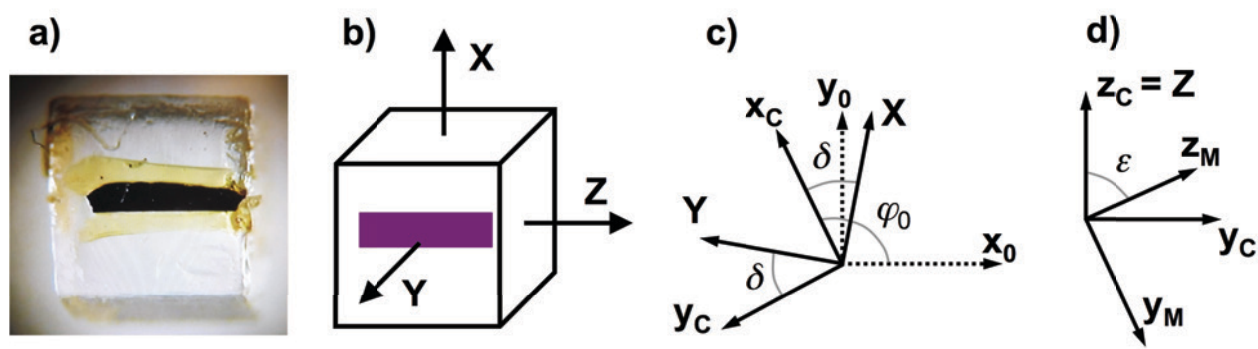

**Figure S3.** a) Picture of a crystal of mounted on a 2x2x2 mm³ cube-shaped holder with a little GE varnish. b) Schematic representation of the sample orientation and definition of the XYZ laboratory axes. c) and d) Representation and relative orientations of the different reference frames used. From Fig. S5, $\varphi_0 \approx 116°$ and $\delta \approx 31°$. Considering the structural information, $\varepsilon = 64.7°$.

Spectra were first acquired for **1$^{VO}$** rotating 360° in 5° steps around each of the three axes defined by the cube-shaped holder X, Y, and Z. The microwave frequency hardly changed over the three series of acquisitions, being 9.8493, 9.8489 and 9.8482 GHz for rotations around X, Y and Z, respectively. Thus $\Delta\nu < 0.3$ MHz so that the frequency can be taken as constant over a full set of measurements since the resolution in magnetic field is 0.2 mT.
By convention, the reference system ($x_0$, $y_0$, $z_0$) is introduced with $z_0 = Z$ and $x_0$ coinciding with the origin of angles for the rotation in the XY plane.
Results for rotations in the XY and YZ planes are shown in Figure 2 of the main text, and reproduced here together with the rotation in the XZ plane in Figure S4.

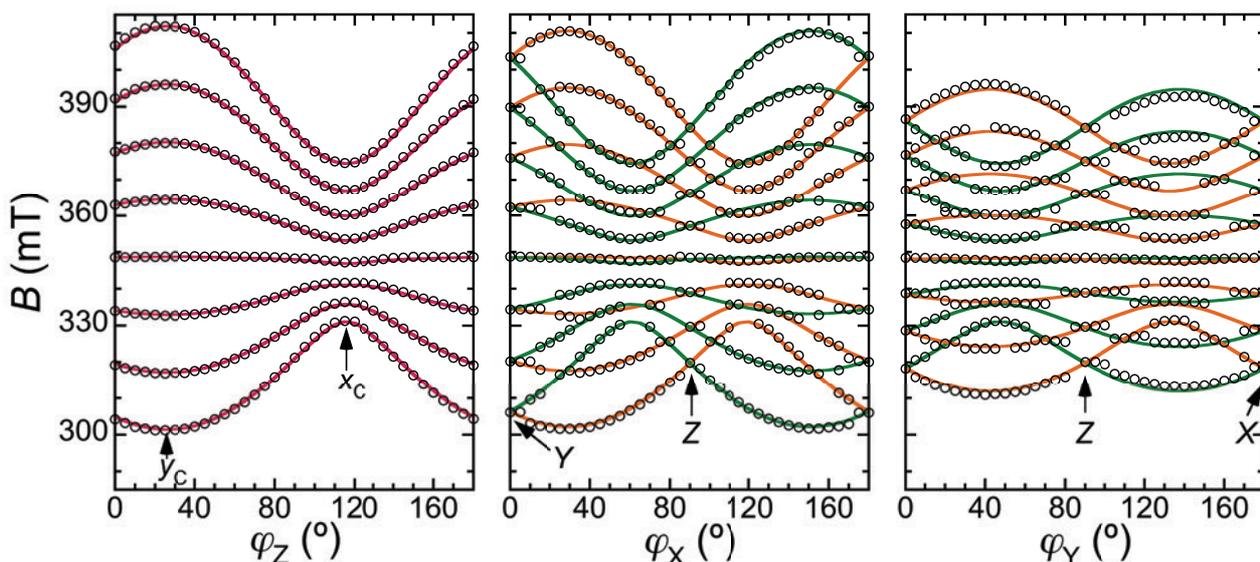

**Figure S4.** Rotational diagrams for a single crystal of **1$^{VO}$** at RT. The empty circles represent the positions of the center of the lines in the CW-EPR spectra upon rotating the crystal around the Z (left), X (middle) and Y (right) axes (see Figure S6b). Full lines are the corresponding positions calculated with the spin Hamiltonian. The contribution from the two magnetically inequivalent orientations of the molecules is represented in different colours (orange and green).

From these experiments, one can already conclude that:

i) for any orientation of the magnetic field perpendicular to the Z axis (i.e. in the XY plane), the 2 orientations of the molecules in the structure are magnetically equivalent. Thus, the rotation around the Z axis provides direct information for the molecule(s) and the symmetry of the rotational diagram exclusively reflects the molecular symmetry.

ii) there are 2 mutually perpendicular directions in the XY plane for which the total splitting of the spectrum is respectively minimal and maximal (see Figure 2 of the main manuscript). These are respectively defined as $x_C$ and $y_C$. Considering the anisotropy of the hyperfine interaction with the $^{51}$V nucleus in vanadyl complexes, $x_C$ will correspond to a direction of the XY plane perpendicular to the $z_M$ of the molecule (in principle $z_M$ corresponds to the V=O direction), and $y_C$ to the projection of $z_M$ on the XY plane (intersection of the $Zz_M$ plane with the XY plane). Then $z_C$ is defined from the $x_C$ and $y_C$ directions, parallel to Z.

iii) if the origin for angles of the rotation in the XY plane $\varphi_Z'$ is taken along $x_C$ (or $y_C$), the rotation diagram is invariant vs. the exchange $\varphi_Z' \rightarrow -\varphi_Z'$.

iv) the comparison of spectra measured with the magnetic field parallel to Y (rotation around X), and X (rotation around Y) with the diagram obtained by rotating around Z allows to define the positions of X and Y axes with respect to the arbitrary origins of the rotational diagram ($x_0$ and $y_0$), and relate these to $x_C$ and $y_C$ (as shown in Figures S5 and S3c).

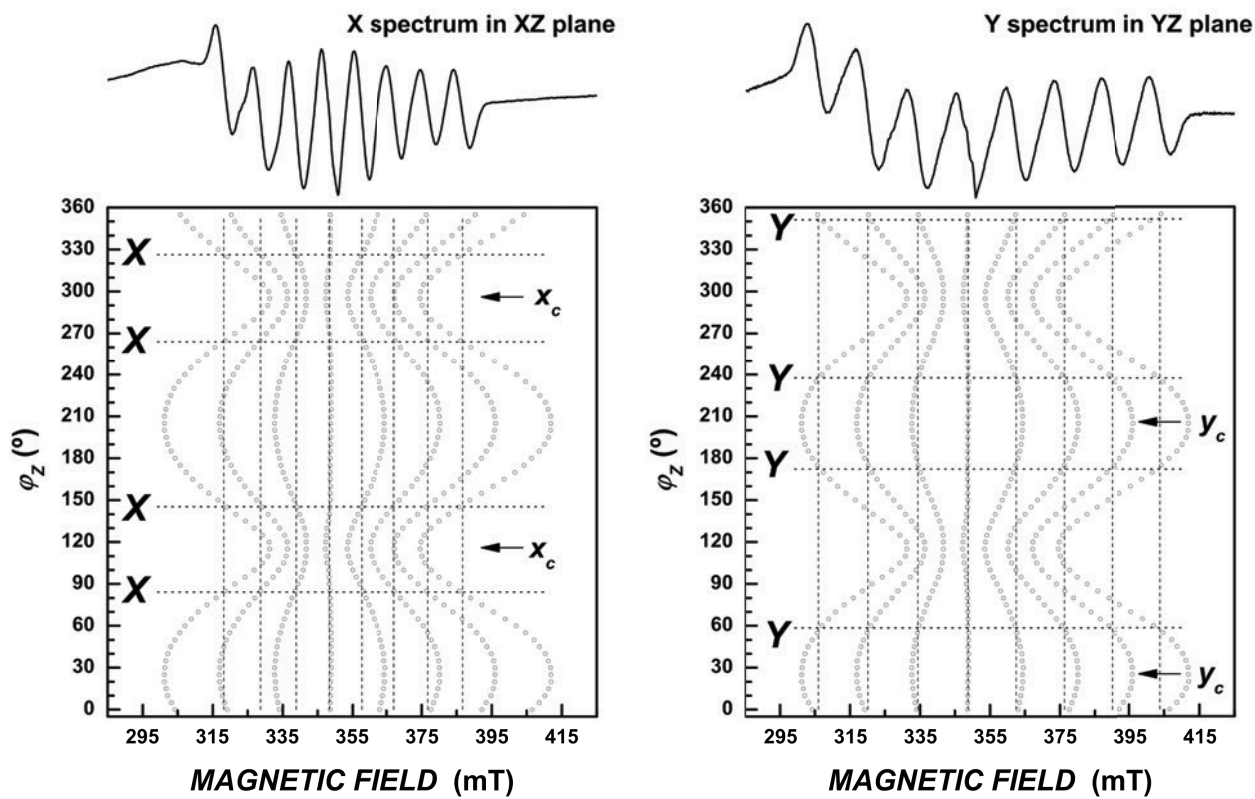

**Figure S5.** CW-EPR spectrum of a single crystal of **1$^{VO}$** with magnetic field along the X (left) and Y (right) axis when rotating in the XZ (left) and YZ (right) plane, allowing to define the x$_C$ and y$_C$ directions, as indicated. A full spectrum is shown on the top, while the positions of the different lines are depicted in the bottom part while rotating.

The symmetry of the rotational diagram around Z indicates that x$_C$ and y$_C$ are either a C$_2$ axis or the intersection of symmetry plane with XY, the latter being equivalent to a binary axis normal to the symmetry plane from a magnetic point of view. This implies that one of the principal directions of the gyromagnetic and hyperfine tensors is along the binary axis, or normal to the symmetry plane, which can be reasoned in terms of the structure of **1$^{VO}$**. Thus, considering that x$_C$ coincides with a principal axis of both gyromagnetic and hyperfine tensors, the other two principal axis will be in the Zy$_C$ plane. Taking into account the molecular structure, one can assume an axial symmetry and define these principal axis with z$_M$ along the V=O moiety and y$_M$ being degenerate with x$_M$, which itself coincides with x$_C$. From the structural parameters, the angle $\varepsilon$ between z$_C$ (or Z) and z$_M$ should be 64.7° (Fig. S3d).

The above definitions then allow calculating the rotational dependence of the EPR spectrum, considering an axial model, for which the position of the different lines will only depend on the angle $\theta$ between the magnetic field and z$_M$. If x' is the intersection of the plane normal to the magnetic field and the plane normal to z$_M$, and defining the reference system (x', y', z') so that z' coincide with z$_M$, the spin Hamilatonian is given by:

$$H = \mu_B B(\sin\theta\, g_\perp S_{x'} + \cos\theta\, g_\parallel S_{z'}) + A_\perp(S_{x'}I_{x'} + S_{y'}I_{y'}) + A_\parallel S_{z'}I_{z'}$$

with the coordinates of unitary vector $\hat{u}$ along z$_M$ in the reference system XYZ given by:

$$\hat{u} = (-\sin\varepsilon\sin\delta,\, \sin\varepsilon\sin\delta,\, \cos\varepsilon)$$

Considering the generic site with $\varepsilon$, $\delta$ and $\varphi_0$ as defined above, one can determine the angle $\theta$ for the different rotations :

1) around X
with the origin of $\varphi_X$ taken as the Y axis (see Fig. S6c), the coordinates of a unitary vector in the direction of the magnetic field in the XYZ reference are [0, $\cos\varphi_X$, $\sin\varphi_X$] and then:

$$\cos\theta(\varphi_X) = \sin\varepsilon\cos\delta\cos\varphi_X + \cos\varepsilon\sin\varphi_X$$

2) around Y
with the origin of $\varphi_X$ taken as the X axis (see Fig. S6c), the coordinates of a unitary vector in the direction of the magnetic field in the XYZ reference are [$\cos\varphi_Y$, 0, $\sin\varphi_Y$] and then:

$$\cos\theta(\varphi_Y) = -\sin\varepsilon\sin\delta\cos\varphi_Y + \cos\varepsilon\sin\varphi_Y$$

3) around Z
with $\varphi_Z$ arbitrary origin taken as x0 (see Fig. S3c), the coordinates of a unitary vector in the direction of the magnetic field in the XYZ reference are [$\cos(\varphi_Z-\varphi_0+\delta)$, $\sin(\varphi_Z-\varphi_0+\delta)$, 0] and then:

$$\cos\theta(\varphi_Z) = \sin\varepsilon\cos(\varphi_Z - \varphi_0)$$

Although there are two magnetically inequivalent orientations of the molecules (say 1 and 2), if $\delta_1 = \delta$ and $\varphi_{01} = \varphi_0$ then $\delta_2 = \delta + \pi$ and $\varphi_{02} = \varphi_0 + \pi$, so that:

$$\cos\theta_1(\varphi_X) = \sin\varepsilon\cos\delta\cos\varphi_X + \cos\varepsilon\sin\varphi_X$$
$$\cos\theta_2(\varphi_X) = -\sin\varepsilon\cos\delta\cos\varphi_X + \cos\varepsilon\sin\varphi_X$$

$$\cos\theta_1(\varphi_Y) = -\sin\varepsilon\sin\delta\cos\varphi_Y + \cos\varepsilon\sin\varphi_Y$$
$$\cos\theta_2(\varphi_Y) = \sin\varepsilon\sin\delta\cos\varphi_Y + \cos\varepsilon\sin\varphi_Y$$

In general the angles $\theta_1$ and $\theta_2$ are different, which results in the observation of two octets in spectra for arbitrary orientations in the YZ and XZ planes. Coalescence of the two occurs for $\varphi_X = \pi/2$ or $\varphi_Y = \pi/2$ (Z direction) as well as for $\varphi_X = 0$ (Y direction) or $\varphi_Y = 0$ (X direction), since then $\theta_1 = \theta_2 = \varepsilon$ or $\theta_1 = \pi - \theta_2$ respectively. The latter situation occurs for any orientation in the XY plane (see Fig. S7).

The spin Hamiltoninan parameters were first estimated considering the spectra corresponding to situation where the magnetic field is oriented along the X, Y, Z, $x_C$ and $y_C$ directions. The spectrum of the polycrystalline sample (Fig. 2 of the main text) was also considered as it provides a good estimation of $g_\parallel$ and $A_\parallel$. On the other hand, the single-crystal spectrum with the field along $x_C$ is dominated by $g_\perp$ and $A_\perp$.
Altogether this gives:

$$g_\parallel = 1.963 \qquad g_\perp = 1.99$$

$$A_\parallel = 475\ MHz \qquad A_\perp = 172\ MHz$$

The excellent agreement between the experimental and calculated spectra is shown in Figure S6 for the five orientations mentioned above, and in Figure S7 for two intermediate orientations. Figure S4 and Fig. 2 of the main text show a good agreement with the positions of the different resonance lines (when sufficiently resolved) for the three rotational diagrams. Figure S8 illustrates the agreement between theory and experimental results using a 2D diagram built from all spectra. For rotations in the XZ and YZ planes, the contribution from the two inequivalent orientations are marked with different colours. There is only a small disagreement for the outer lines measured in the XZ plane, which we ascribe to a small misalignement of the cube-shaped holder as it was placed onto the main sample holder.

The same procedure was applied for **1$^{VO}_{3\%}$**, for which the data obtained are virtually identical, as the same spin Hamiltonian parameters are able to reproduce the experimental spectra and the positions of all resonance lines (see Figure 5 of the main text and Figure S9). There is however an additional isotropic signal at g = 2.01 in all spectra, which we ascribe to a Ti(III) impurity. Some disagreement is also observed between the experimental and calculated angle dependence for the rotation in the XZ plane, which we again ascribe to a small error in the alignment of the cube-shaped holder.

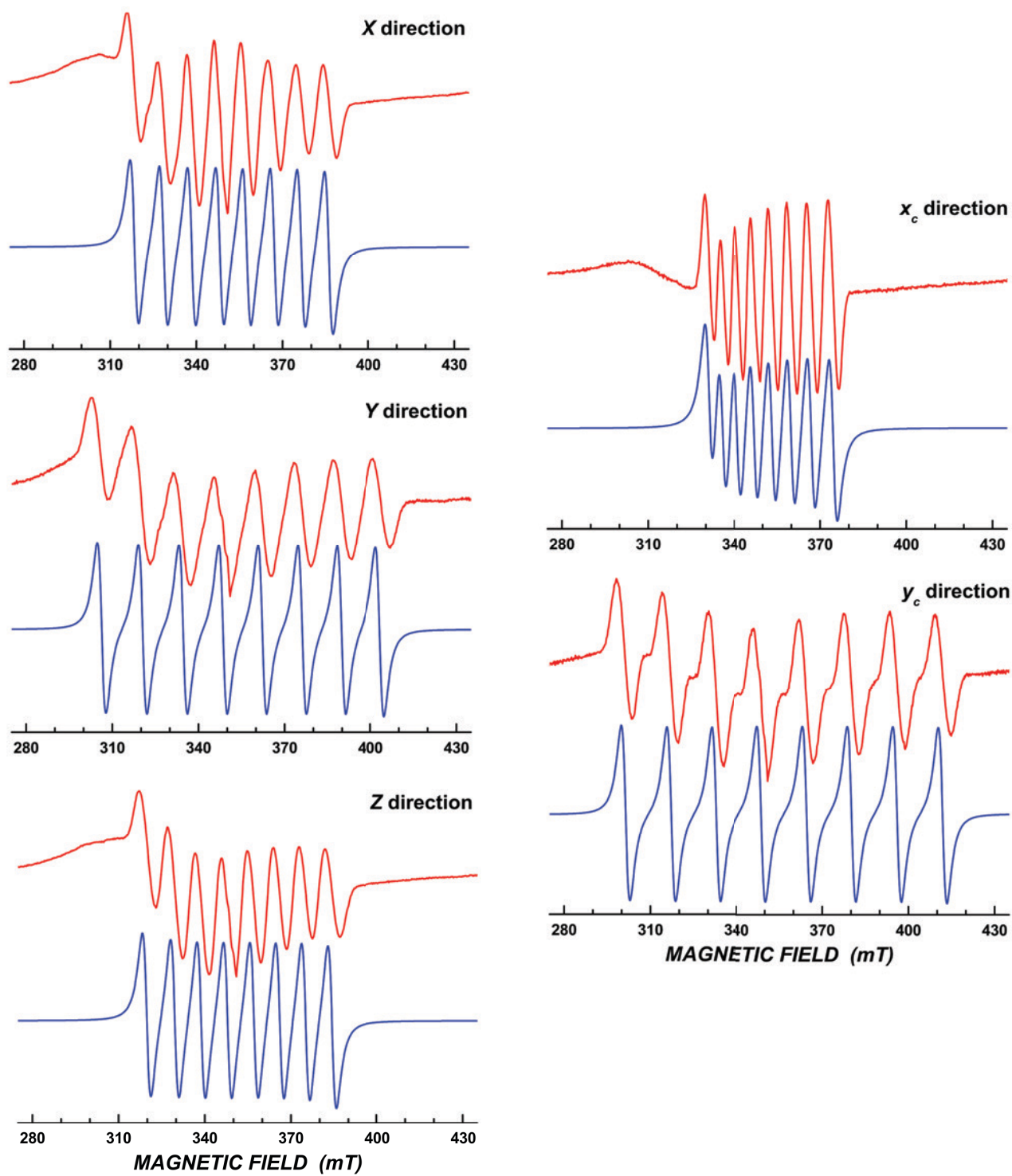

**Figure S6.** Comparison of experimental (red lines) and calculated (blue lines) CW-EPR spectra of a single crystal of **1^VO** with the magnetic field along the indicated directions. Calculated spectra were obtained using a Lorentzian lineshape with a 3 mT isotropic width.

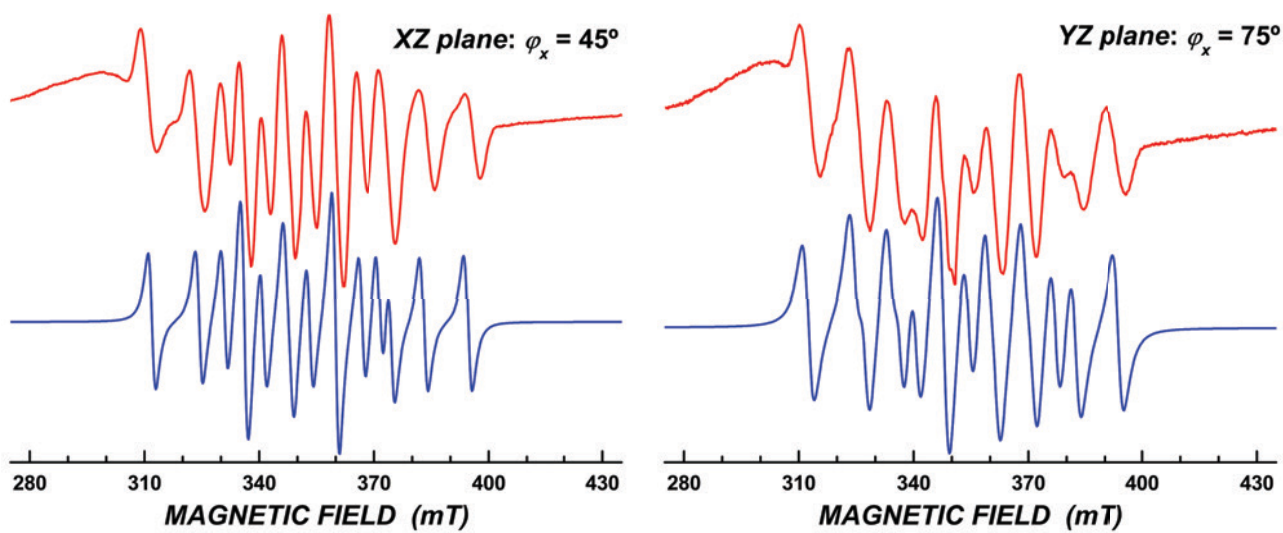

**Figure S7.** Comparison of experimental (red lines) and calculated (blue lines) CW-EPR spectra of a single crystal of **1**$^{VO}$ with the magnetic field along the indicated directions. Calculated spectra were obtained using a Lorentzian lineshape with a 3 mT isotropic width.

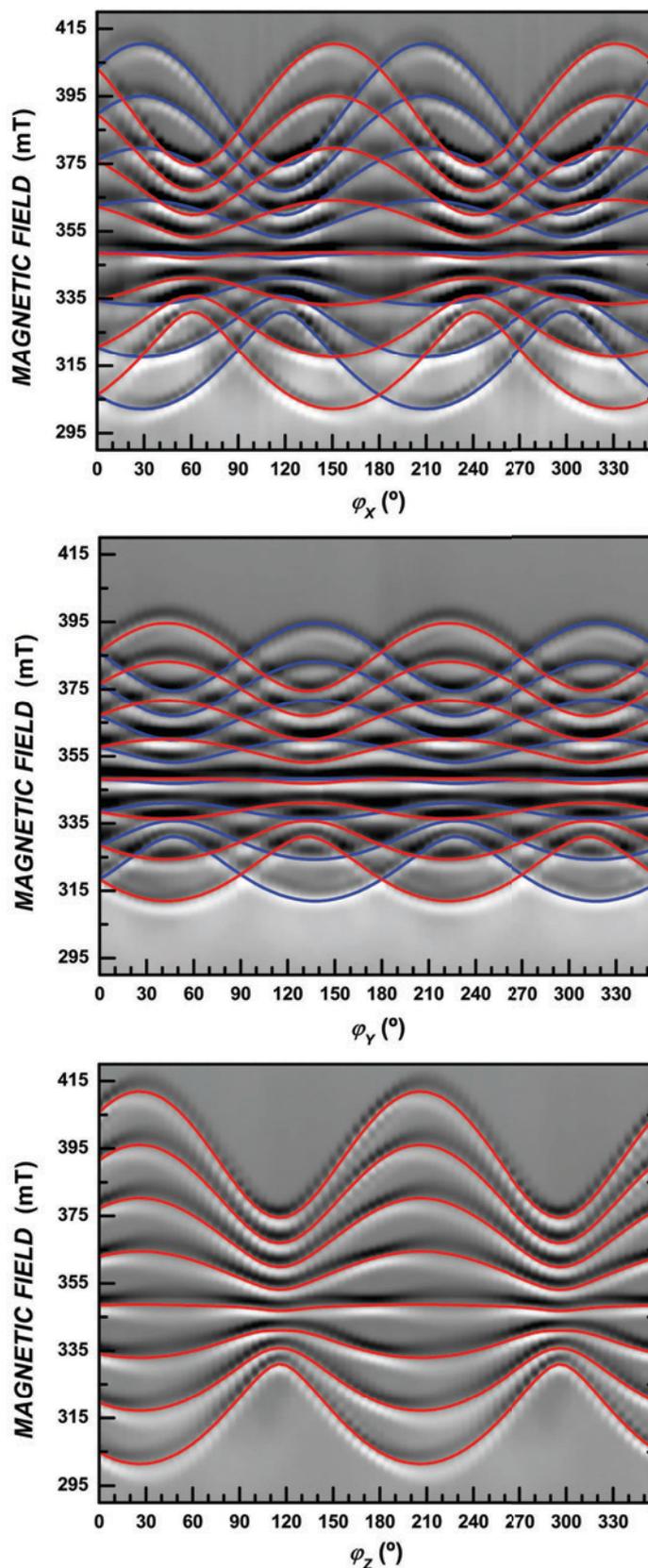

**Figure S8.** Full rotational diagrams for a single crystal of **1**$^{VO}$ at RT. Lines are the expected angle dependence calculated with the spin Hamiltonian. The contribution from the two magnetically inequivalent orientations of the molecules is represented in different colours (red and blue).

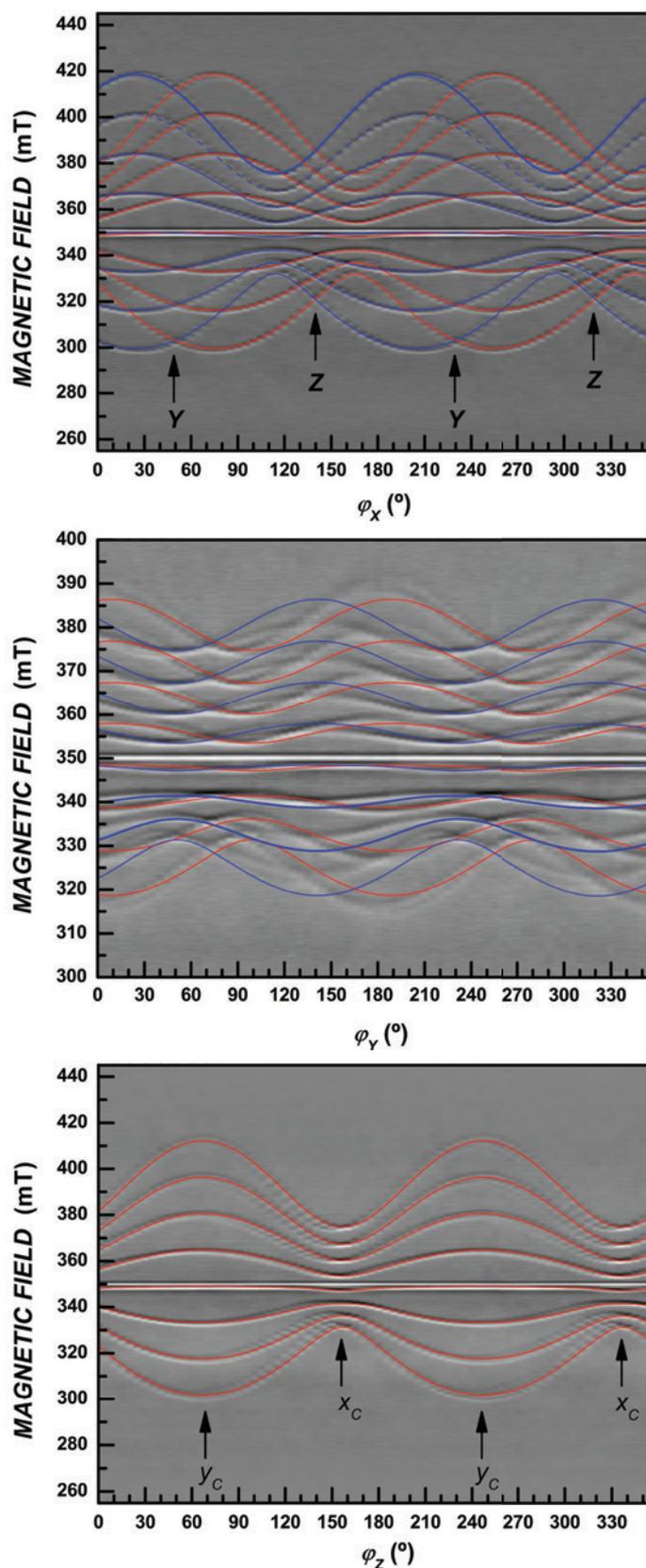

**Figure S9.** Full rotational diagrams for a single crystal of **1**$^{VO}_{3\%}$ at RT. Lines are the expected angle dependence calculated with the spin Hamiltonian. The contribution from the two magnetically inequivalent orientations of the molecules is represented in different colours (red and blue). Arrows indicate the Y, Z, $x_C$ and $y_C$ axes as defined above. The isotropic signal at g = 2.01 present in all spectra is ascribed to a Ti(III) impurity.

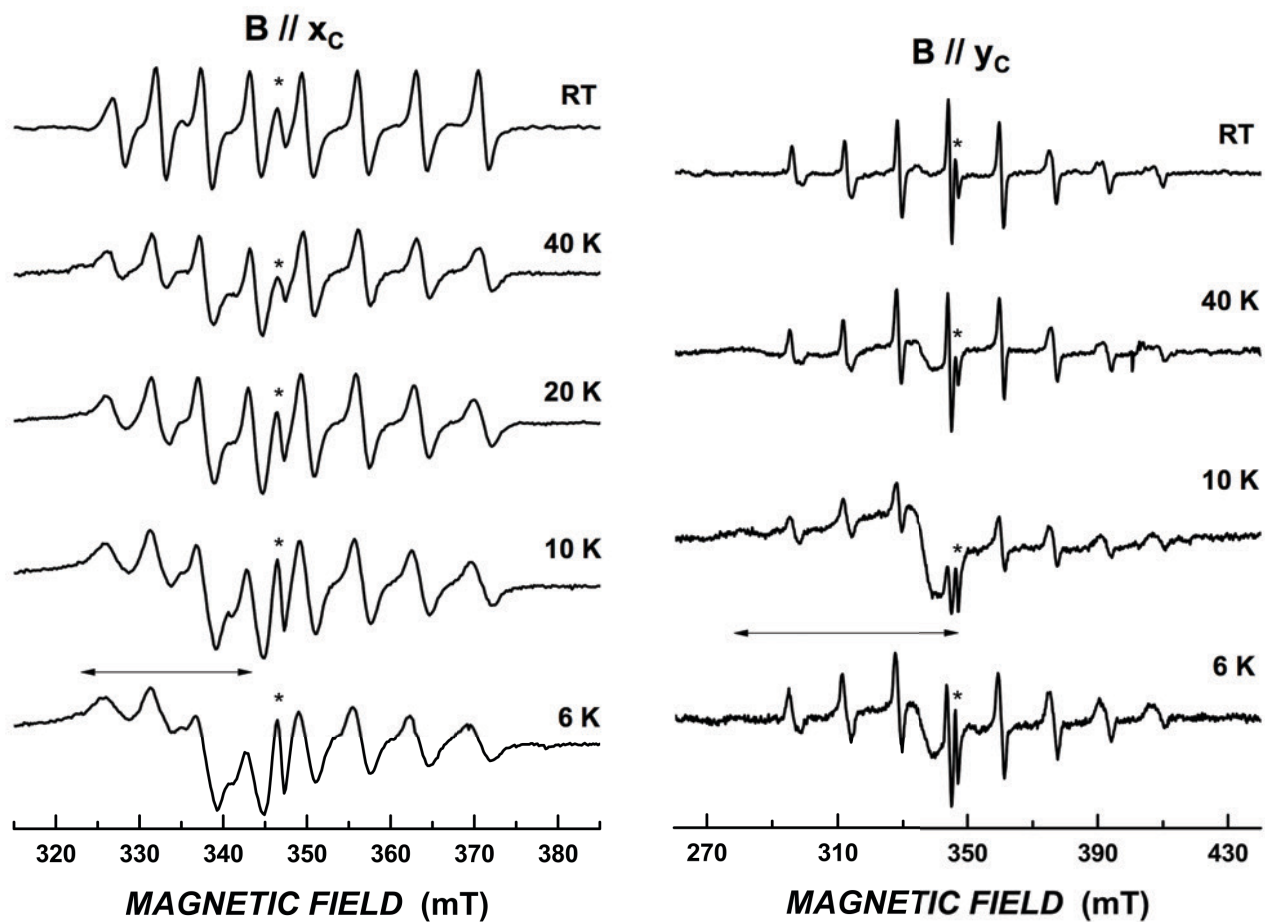

**Figure S10.** CW-EPR spectra of a single-crystal of **1$^{VO}_{3\%}$** measured at different decreasing temperatures and for a magnetic field applied along the $x_C$ (left) and $y_C$ (right) crystal axes. A broad signal covering the 290-350 mT field range arises from Cu(II) ions on the surface of the cavity. The isotropic signal at g = 2.01 present in all spectra and indicated by a star is ascribed to a Ti(III) impurity. A broad signal arising from Cu(II) ions on the surface of the cavity is indicated by horizontal double arrows.

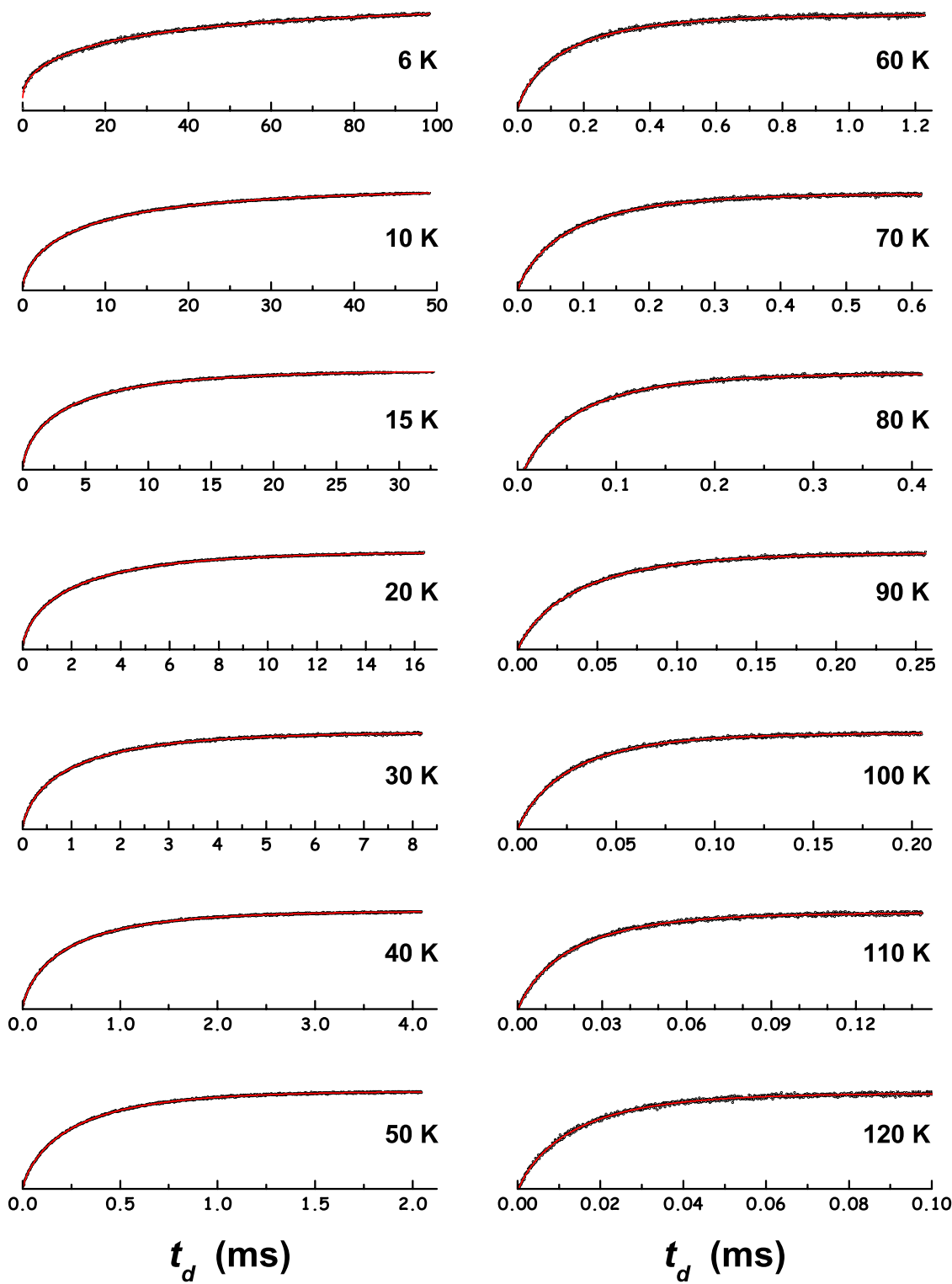

**Figure S11.** ESE detected Inversion Recovery ($\tau$ = 200 ns) as a function of delay time, $t_d$, for the frozen solution **1**$^{VO}{}_{sol}$ at the indicated temperatures and $B$ = 346.6 mT (circles). Red lines correspond to a stretched exponential $t_d$ dependence modelled by:

$$y(t_d) = y_\infty - y_0 e^{-(t_d/\beta T_1)^\beta}$$

The derived values of $T_1$ are depicted in Figure 4 of the main article (see also Table S3).

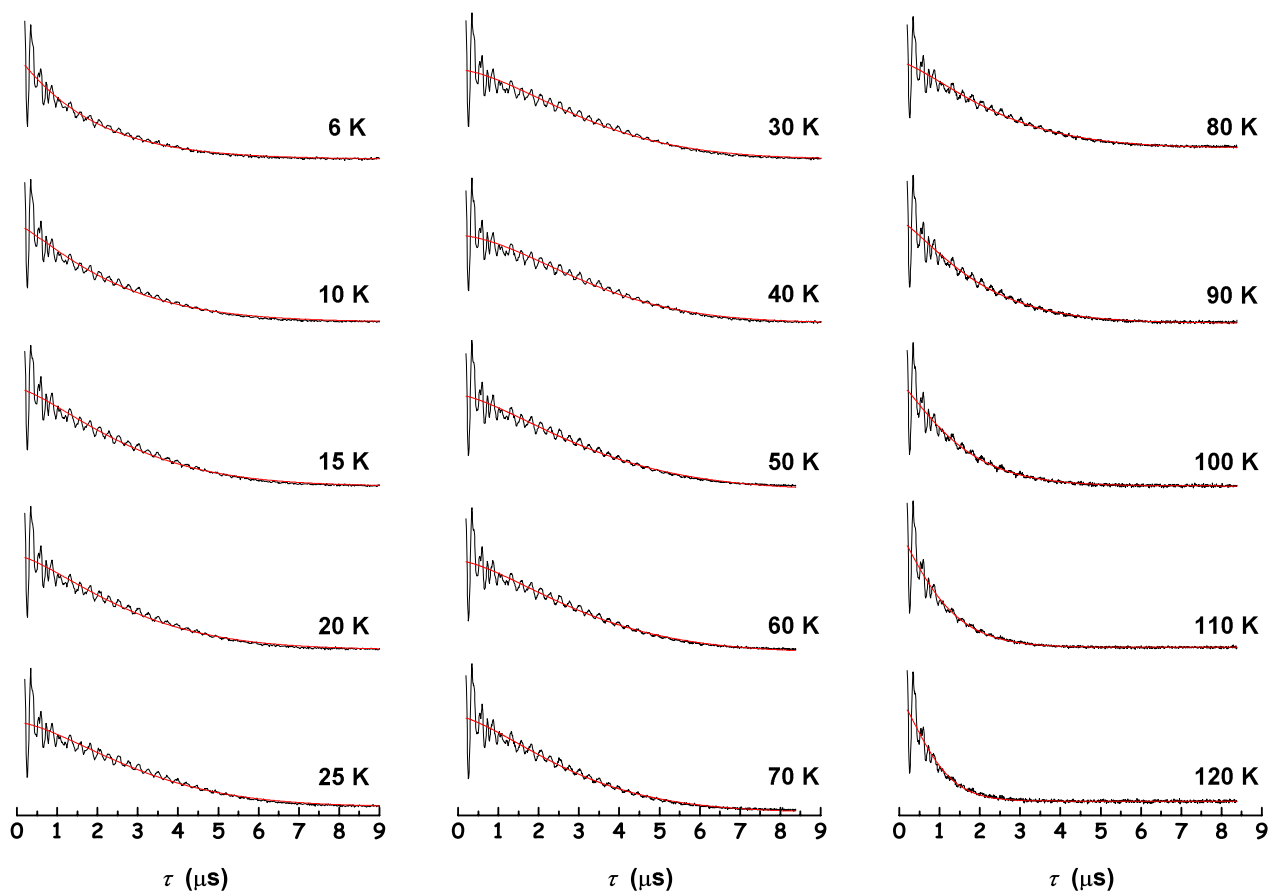

**Figure S12.** 2p ESE intensity as a function of inter-pulse interval, τ, measured on **1**$^{VO}_{sol}$ at the indicated temperatures and *B* = 346.6 mT. Red lines are least-square fits with a stretched exponential:

$$y(\tau) = y_0 + A_{2p} e^{-\left(2\tau/T_2\right)^{\beta}}$$

The values of $T_2$ obtained from these fits are depicted in Figure 3 of the main text (see also Table S3).

At all temperatures, the derived values of *β* are > 1 (in the range 1.05-1.70), which impedes to interpret it as a distribution of relaxation times. Values of *β* > 1 have been associated with spin diffusion processes induced by either the dipolar interaction or the nucleus-nucleus interaction which modulates the spin dephasing through the hyperfine interaction. Lacking additional information, we prefer to avoid giving any interpretation. We trust the derived phase coherence times are robust, as no significant differences result from using the present stretched exponential model or a simple exponential with or without a low frequency modulation (see caption of Figure S13 below).

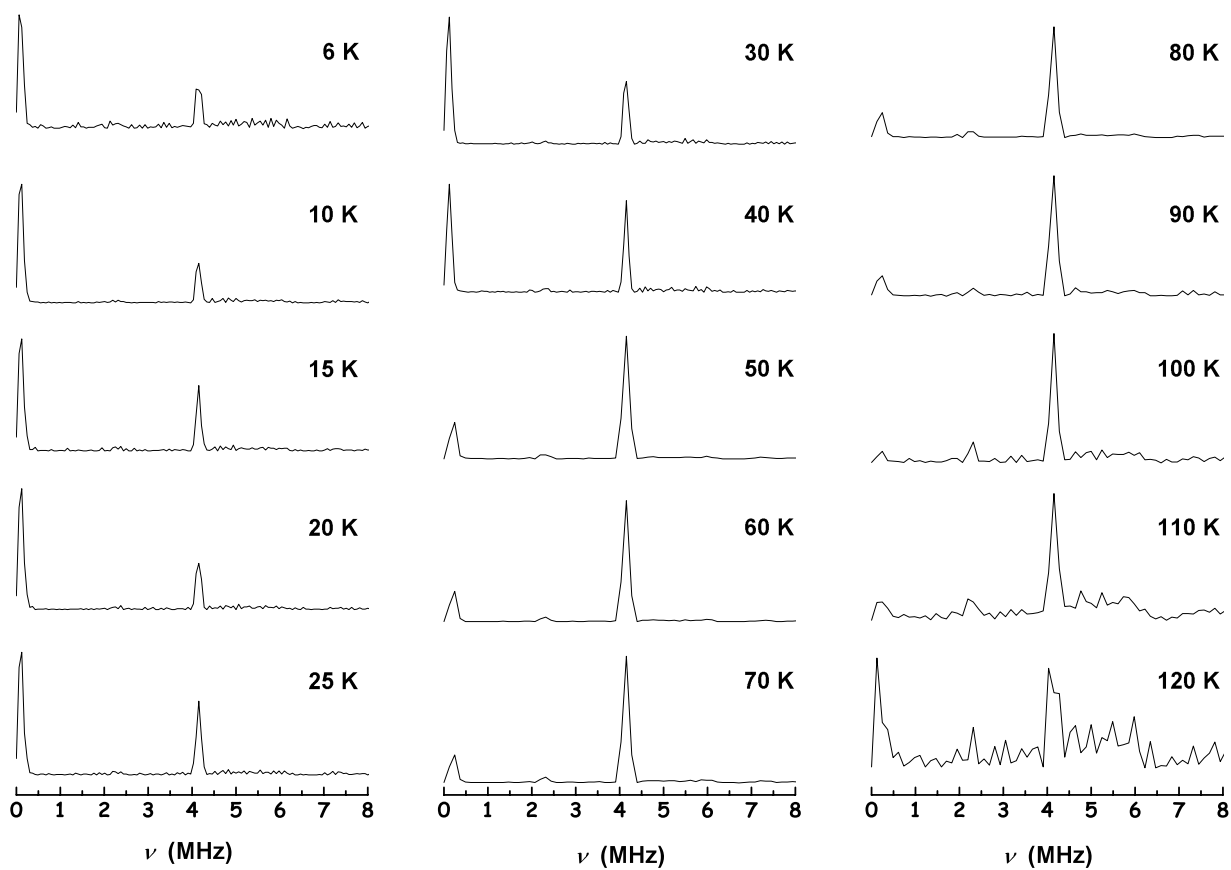

**Figure S13.** Power spectra of the modulation in the 2p ESE decay detected for $\mathbf{1^{VO}_{sol}}$ at the indicated temperatures and $B$ = 346.6 mT.

The observed low frequency contribution may be due to experimental variations of the background or small deviations from an exponential decay. It can be taken into account by considering an exponential decay with low frequency modulation using:

$$y(\tau) = y_0 + A\{e^{-2\tau/T_2} + k e^{-2\tau/T_D} \cos(2\pi\tau\nu) + \varphi_0\}$$

with the frequency $\nu$ in the range 0.10-0.14 MHz.

While the simulation of the 2p ESE decay is good and the low frequency contribution is then mostly removed from the resulting power spectra, the times $T_2$ and $T_D$ obtained in this way show significantly larger errors, especially at temperatures above 50 K. Since these are similar as those obtained with the stretched exponential model, we have preferred to keep the later.

**Table S3.** Parameters that provide the best fits to ESE detected Inversion Recovery signals and the 2p-ESE decay for **1$^{VO}_{sol}$** measured at 346.6 mT and variable temperature.

| T (K) | T$_1$ (μs) | β | T$_2$ (μs) | β |
|---|---|---|---|---|
| 6 | 52573 ± 1521 | 0.534 ± 0.007 | 3.76 ± 0.20 | 1.05 ± 0.04 |
| 10 | 13050 ± 61 | 0.603 ± 0.003 | 5.17 ± 0.26 | 1.23 ± 0.04 |
| 15 | 5862 ± 19 | 0.639 ± 0.002 | 6.06 ± 0.28 | 1.41 ± 0.04 |
| 20 | 3168 ± 10 | 0.678 ± 0.002 | 6.16 ± 0.30 | 1.41 ± 0.04 |
| 25 | | | 6.82 ± 0.34 | 1.52 ± 0.04 |
| 30 | 1500 ± 6 | 0.679 ± 0.003 | 7.27 ± 0.36 | 1.60 ± 0.04 |
| 40 | 679 ± 2 | 0.720 ± 0.002 | 7.46 ± 0.36 | 1.70 ± 0.04 |
| 50 | 349 ± 1 | 0.752 ± 0.002 | 7.42 ± 0.69 | 1.46 ± 0.06 |
| 60 | 197.4 ± 1.2 | 0.766 ± 0.006 | 7.09 ± 0.61 | 1.50 ± 0.06 |
| 70 | 104.2 ± 0.5 | 0.805 ± 0.005 | 6.35 ± 0.50 | 1.46 ± 0.06 |
| 80 | 68.5 ± 0.3 | 0.803 ± 0.004 | 5.27 ± 0.38 | 1.38 ± 0.05 |
| 90 | 47.7 ± 0.2 | 0.825 ± 0.004 | 4.13 ± 0.28 | 1.28 ± 0.05 |
| 100 | 33.0 ± 0.2 | 0.841 ± 0.005 | 3.13 ± 0.21 | 1.23 ± 0.05 |
| 110 | 23.6 ± 0.1 | 0.829 ± 0.005 | 2.27 ± 0.16 | 1.21 ± 0.06 |
| 120 | 16.7 ± 0.1 | 0.829 ± 0.007 | 1.92 ± 0.15 | 1.41 ± 0.07 |

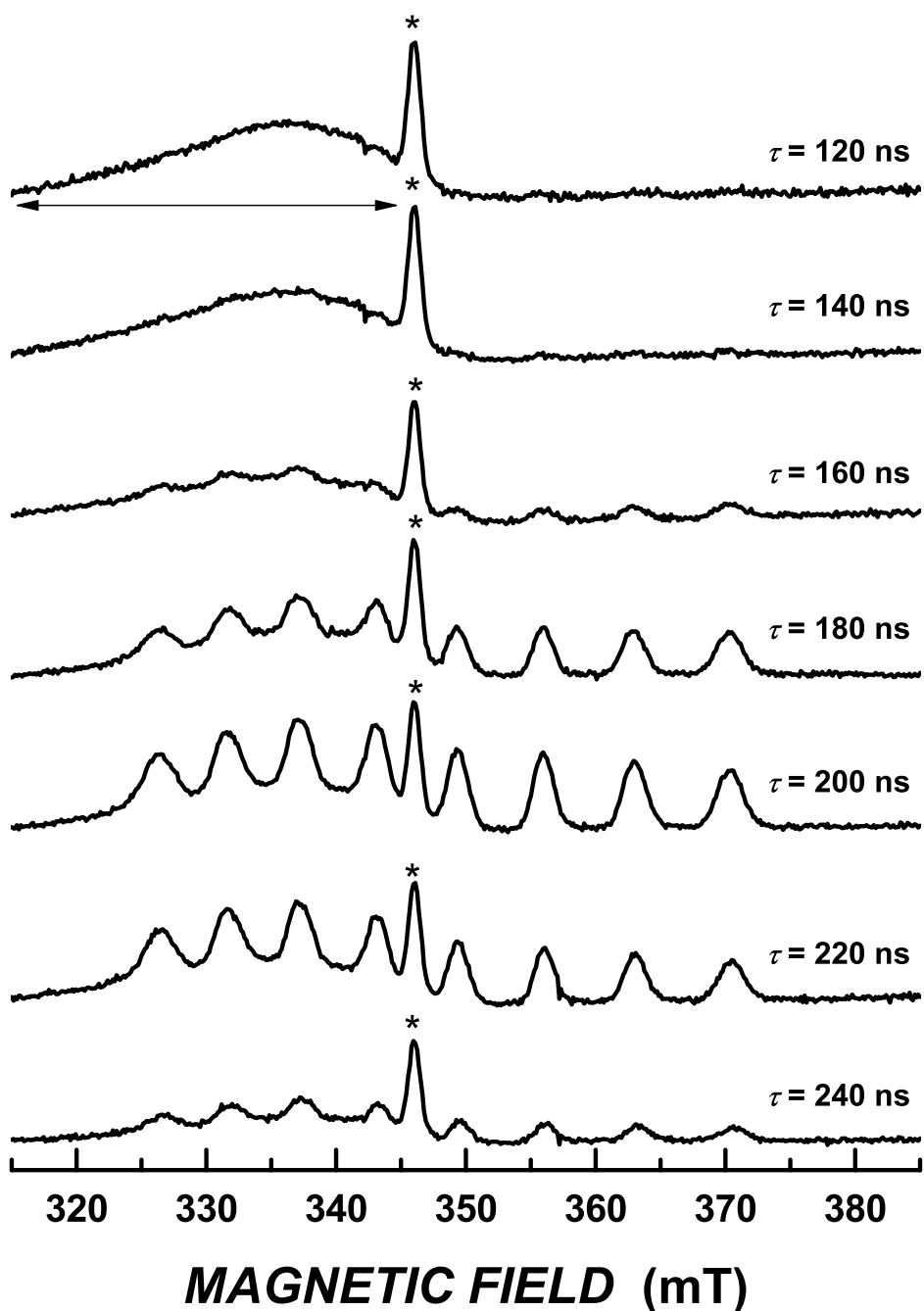

**Figure S14.** ESE-detected EPR spectra measured on a single-crystal of $1^{VO}_{3\%}$ at 6 K and for variable $\tau$, with the magnetic field applied along the $x_C$ crystal axis. A broad signal covering the 290-350 mT field range is detected at all $\tau$ arising from Cu(II) ions on the surface of the cavity. The isotropic signal at g = 2.01 present in all spectra and marked with a star is ascribed to a Ti(III) impurity.

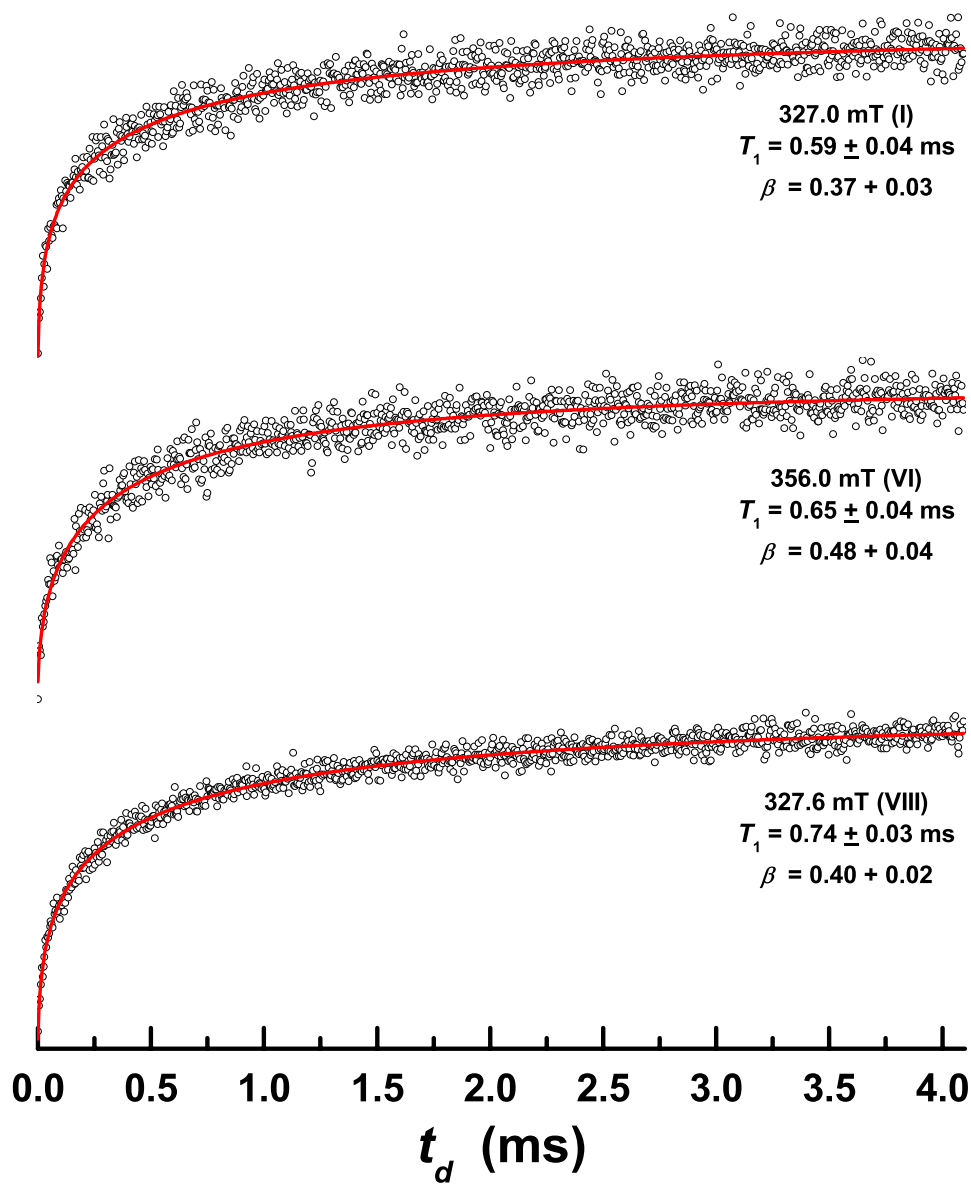

**Figure S15.** ESE detected Inversion Recovery ($\tau$ = 200 ns) measured as a function of delay time, $t_d$, on **1$^{VO}_{3\%}$** at 6 K and for the indicated magnetic fields applied along the $x_C$ direction (circles). Red lines correspond to a stretched exponential $t_d$ dependence modelled by:

$$y(t_d) = y_\infty - y_0 e^{-(t_d/\beta T_1)^\beta}$$

and the indicated best-fit parameters $T_1$ and $\beta$. $T_1$ values are depicted plotted in Figure 5 of the main article.

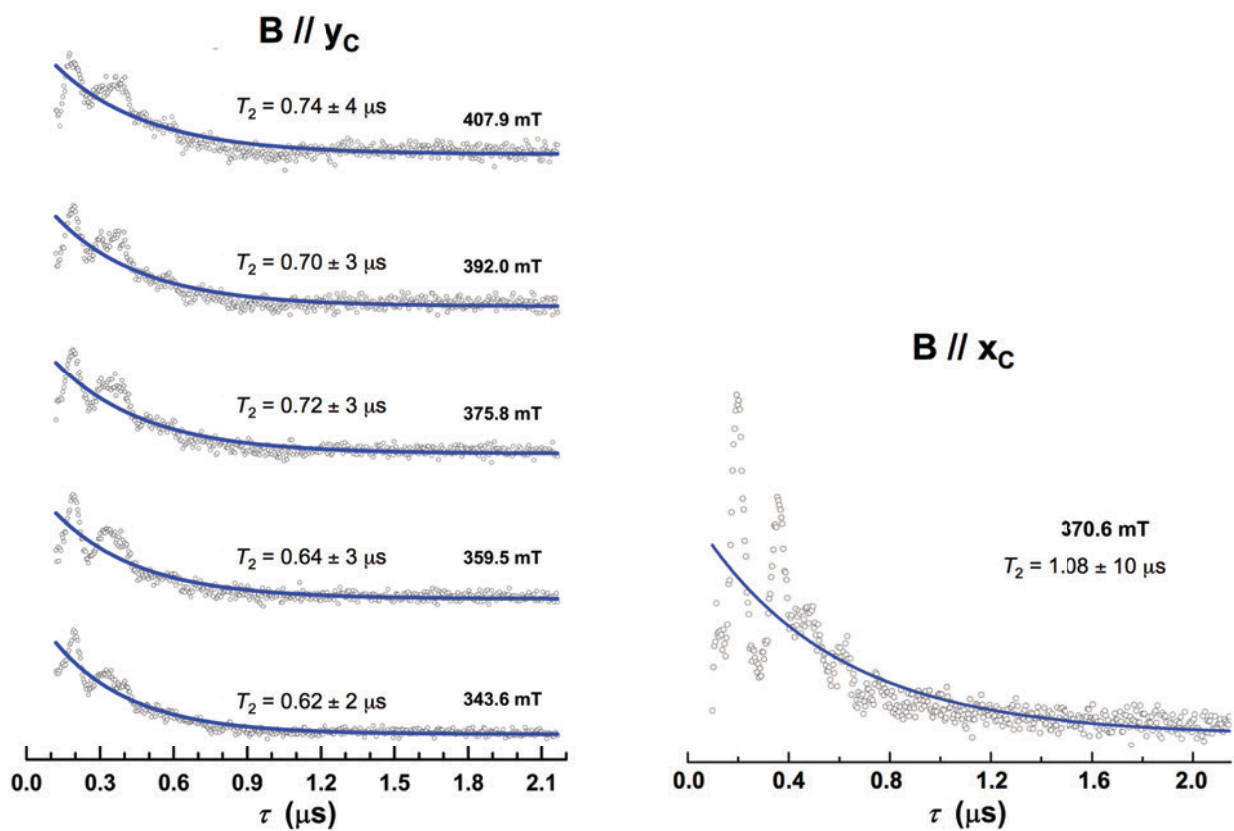

**Figure S16.** 2p ESE intensity measured as a function of inter-pulse interval, τ, on **1$^{VO}_{3\%}$** at 6 K and for the indicated magnetic field orientations and values (circles). Blue lines are least-squares fits with an exponential decay:

$$y(\tau) = y_0 + A_{2p}e^{-2\tau/T_2}$$

and the indicated values of $T_2$, which are depicted in Figure 4 of the main text.

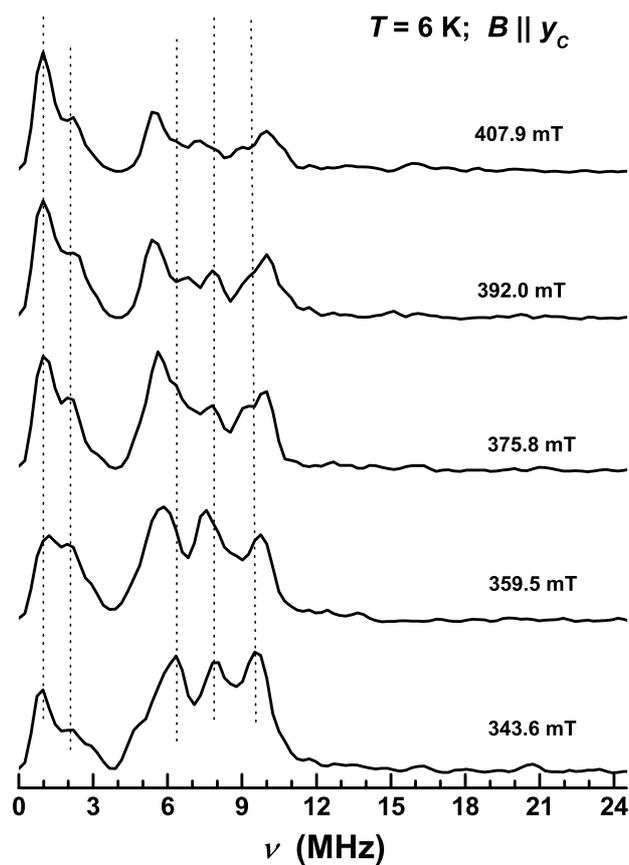

**Figure S17.** Power spectra of the modulation in the 2p ESE decay detected for **1$^{VO}_{3\%}$** at $T$ = 6 K and the indicated magnetic fields applied parallel to the $y_C$ axis. Vertical dash lines are guides to the eye indicating the position of the main frequencies of modulation at the lowest field.

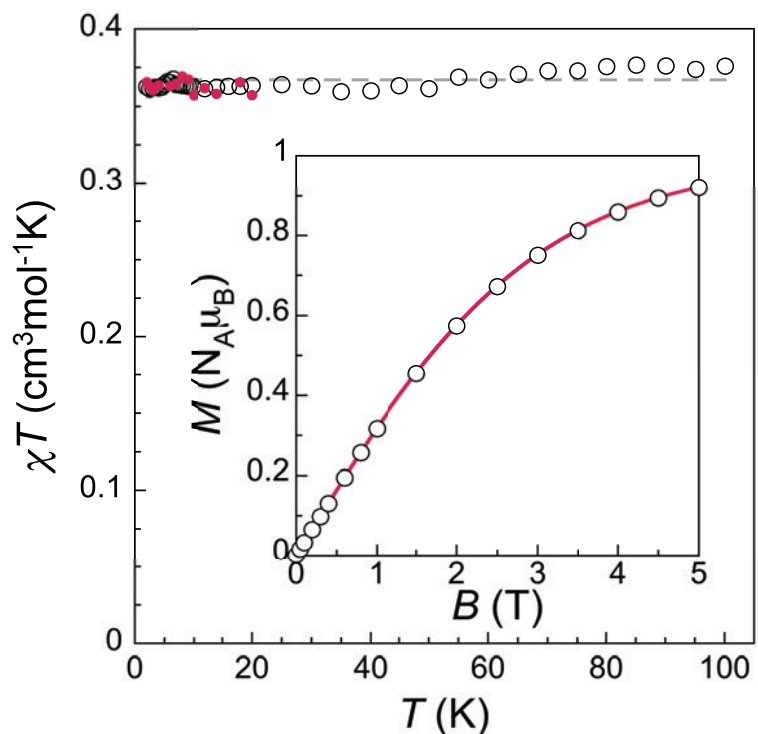

**Figure S18.** Temperature dependence of the $\chi T$ product of **1$^{VO}$** derived from magnetization measurements at 0.01 T (open symbols) and from zero-field *ac* susceptibility measurements at 10 Hz (solid red symbols). The dashed grey line represents the Curie law for $C$ = 0.367 cm$^3$mol$^{-1}$. Inset: magnetization isotherm of **1$^{VO}$** measured at $T$ = 2 K with the corresponding Brillouin function for $S$ = 1/2 and g = 1.98 shown as a red solid line. Both sets of data are thus in excellent agreement with the EPR data from which $g_{\parallel}$ = 1.963 and $g_{\perp}$ = 1.99 were derived.

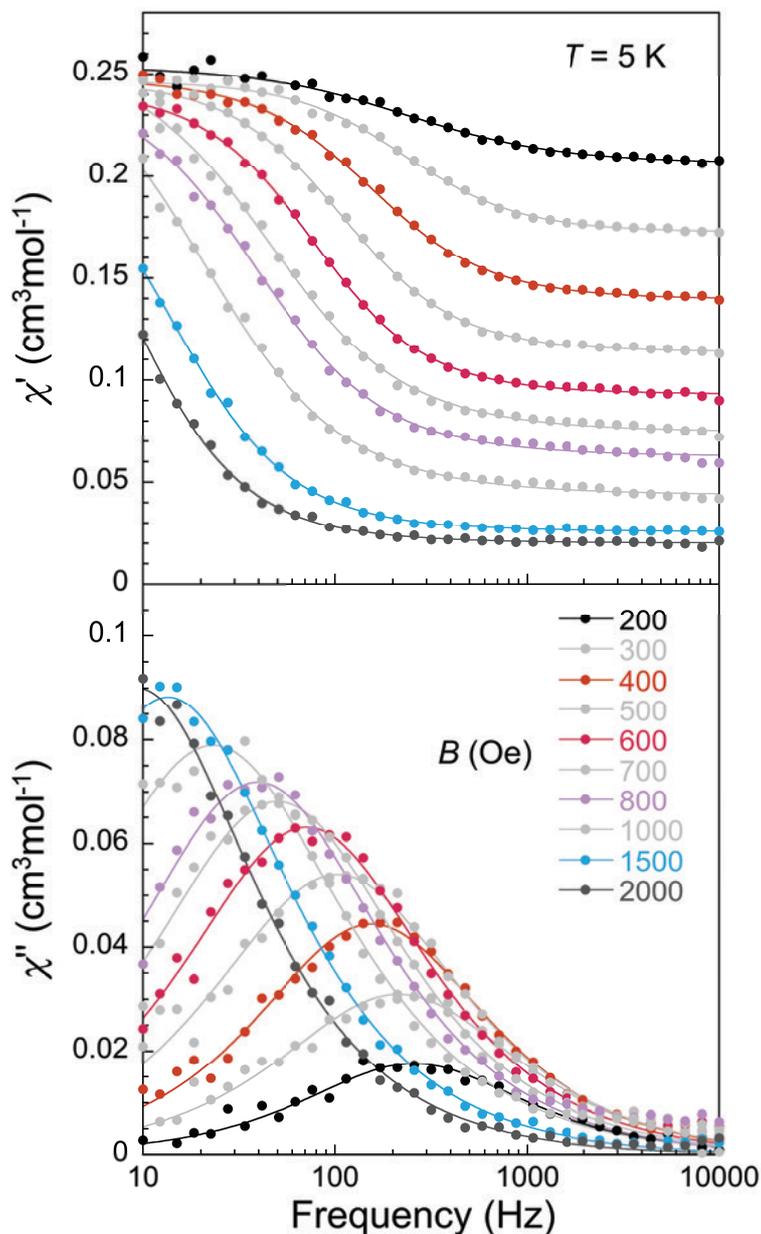

**Figure S19.** Frequency dependence of the in-phase (top) and out-of-phase (bottom) *ac* magnetic susceptibility components of **1**$^{\text{VO}}$ measured at 5 K and for different *dc* applied magnetic fields, as indicated. Lines are least-square fits to the Cole-Cole expressions for $\chi'$ and $\chi''$ that allow determining the characteristic spin relaxation time $\tau$:

$$\chi'(\omega) = \chi_S + (\chi_T - \chi_S)\frac{1 + (\omega\tau)^\beta \cos\left(\frac{\pi\beta}{2}\right)}{1 + 2(\omega\tau)^\beta \cos\left(\frac{\pi\beta}{2}\right) + (\omega\tau)^{2\beta}}$$

$$\chi''(\omega) = (\chi_T - \chi_S)\frac{(\omega\tau)^\beta \sin\left(\frac{\pi\beta}{2}\right)}{1 + 2(\omega\tau)^\beta \cos\left(\frac{\pi\beta}{2}\right) + (\omega\tau)^{2\beta}}$$

in which $\omega$ is the angular frequency, $\chi_T$ the isothermal susceptibility, $\chi_S$ the adiabatic susceptibility and $\beta$ describes a limited distribution of relaxation times (0.8-0.9 range). For a $S$ = 1/2 spin system, $\tau$ coincides with the spin-lattice relaxation time $T_1$.

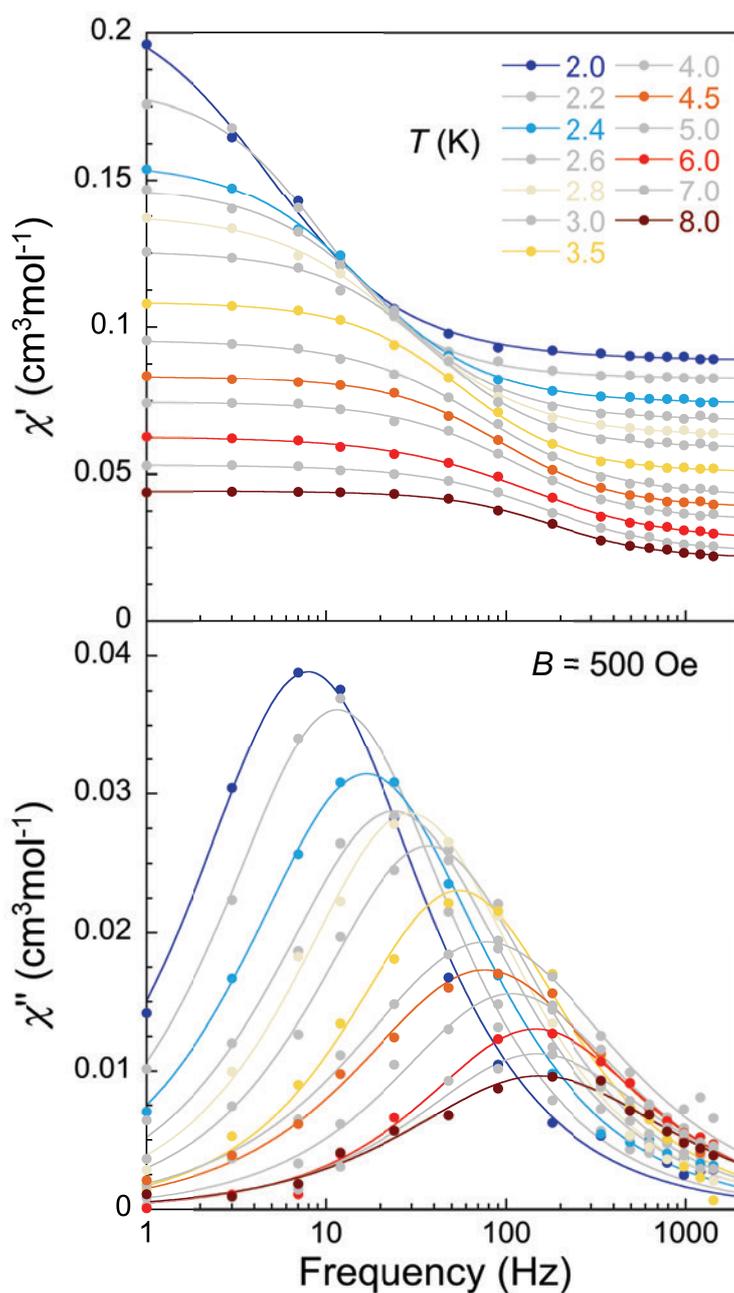

**Figure S20.** Frequency dependence of the in-phase (top) and out-of-phase (bottom) *ac* magnetic susceptibility components of **1$^{VO}$** measured at 500 Oe and for variable *T*, as indicated. Lines are least-square fits to the Cole-Cole expressions for $\chi'$ and $\chi''$ (see Fig S18). Data obtained with a commercial magnetometer using a SQUID sensor.

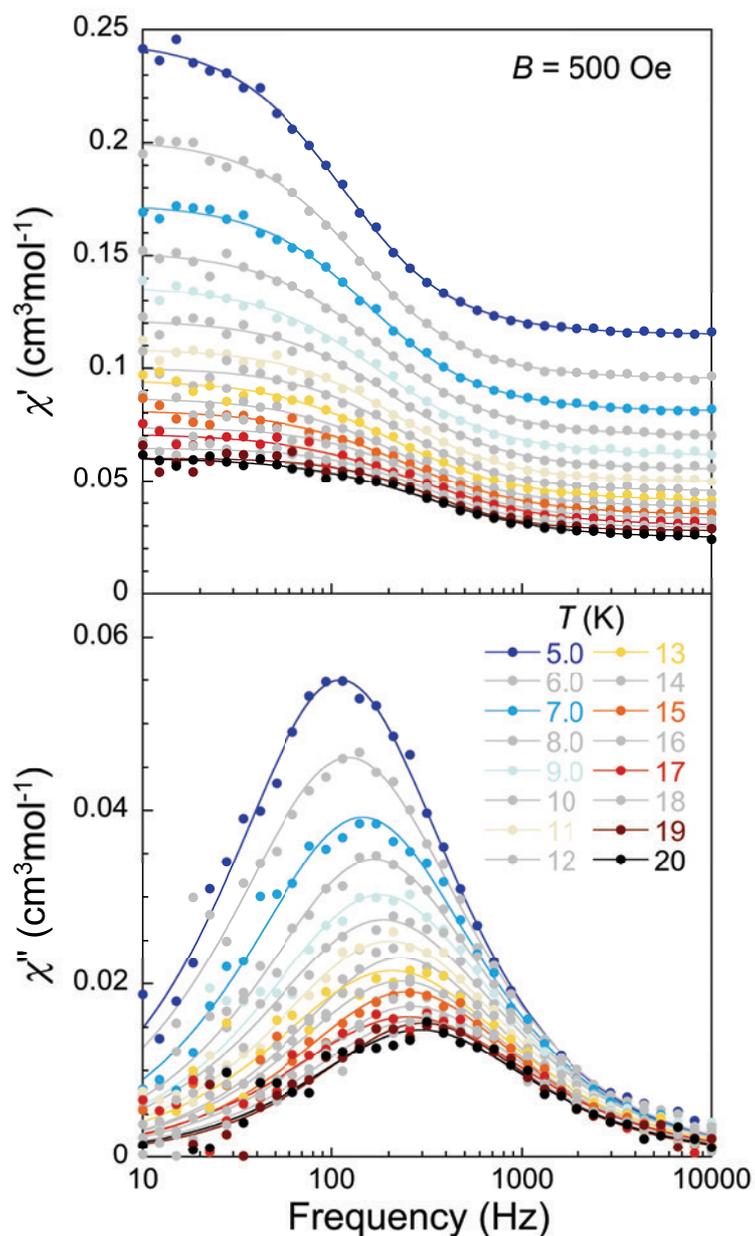

**Figure S21.** Frequency dependence of the in-phase (top) and out-of-phase (bottom) *ac* magnetic susceptibility components of **1$^{VO}$** measured at 500 Oe and for variable *T*, as indicated. Lines are least-square fits to the Cole-Cole expressions for $\chi'$ and $\chi''$ (see Fig S18). Data obtained with the ACMS option of a commercial PPMS set-up.

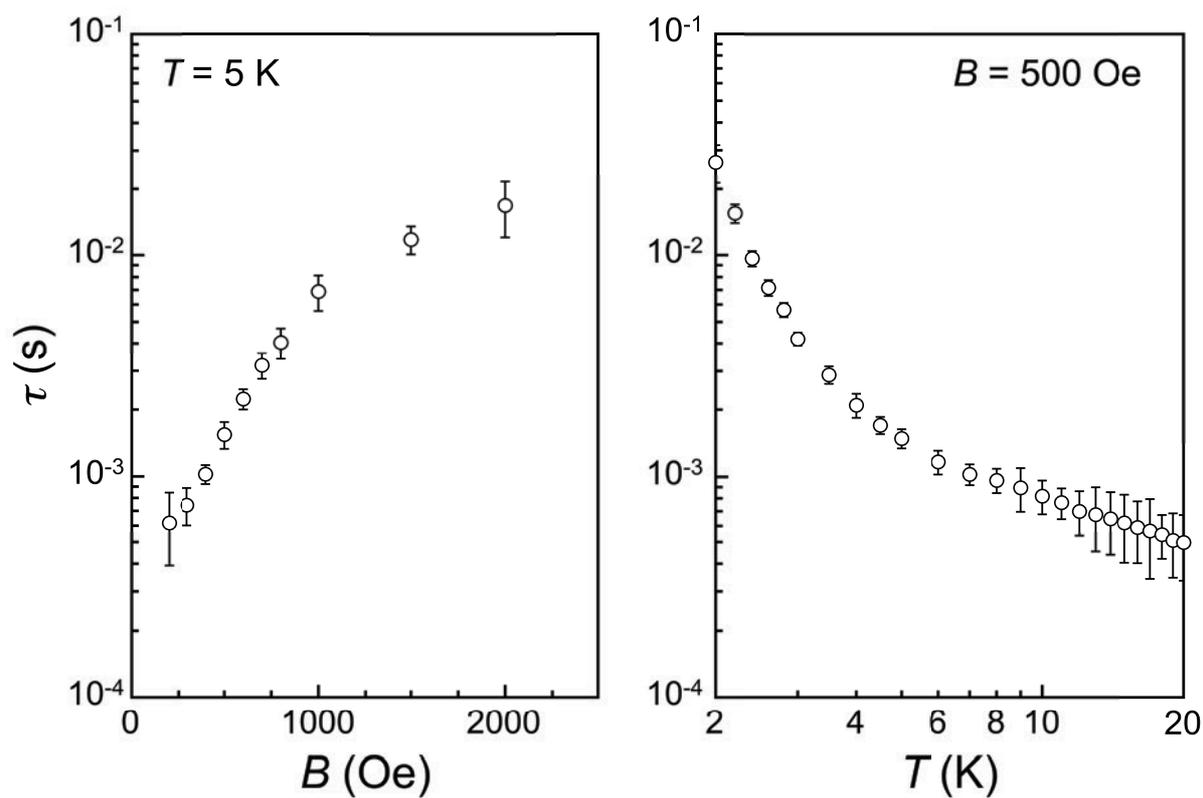

**Figure S22.** Spin-lattice relaxation time of **1^VO** determined from the fit of ac magnetic susceptibility data (Figs. S18-S20).

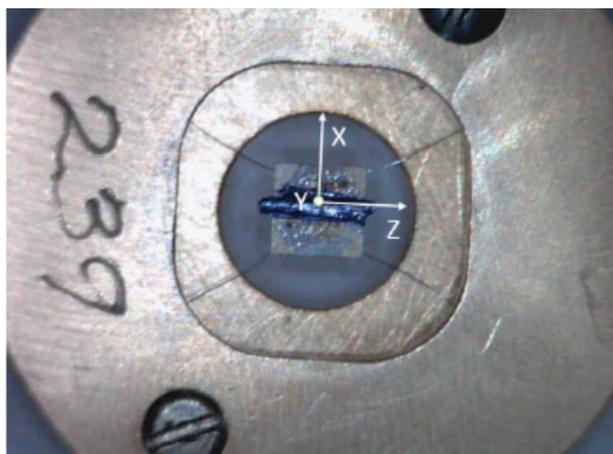

**Figure S23.** Image of a calorimeter platform hosting a single crystal of **1**$^{VO}$. The arrows show the laboratory frame. The magnetic field was applied along Z.

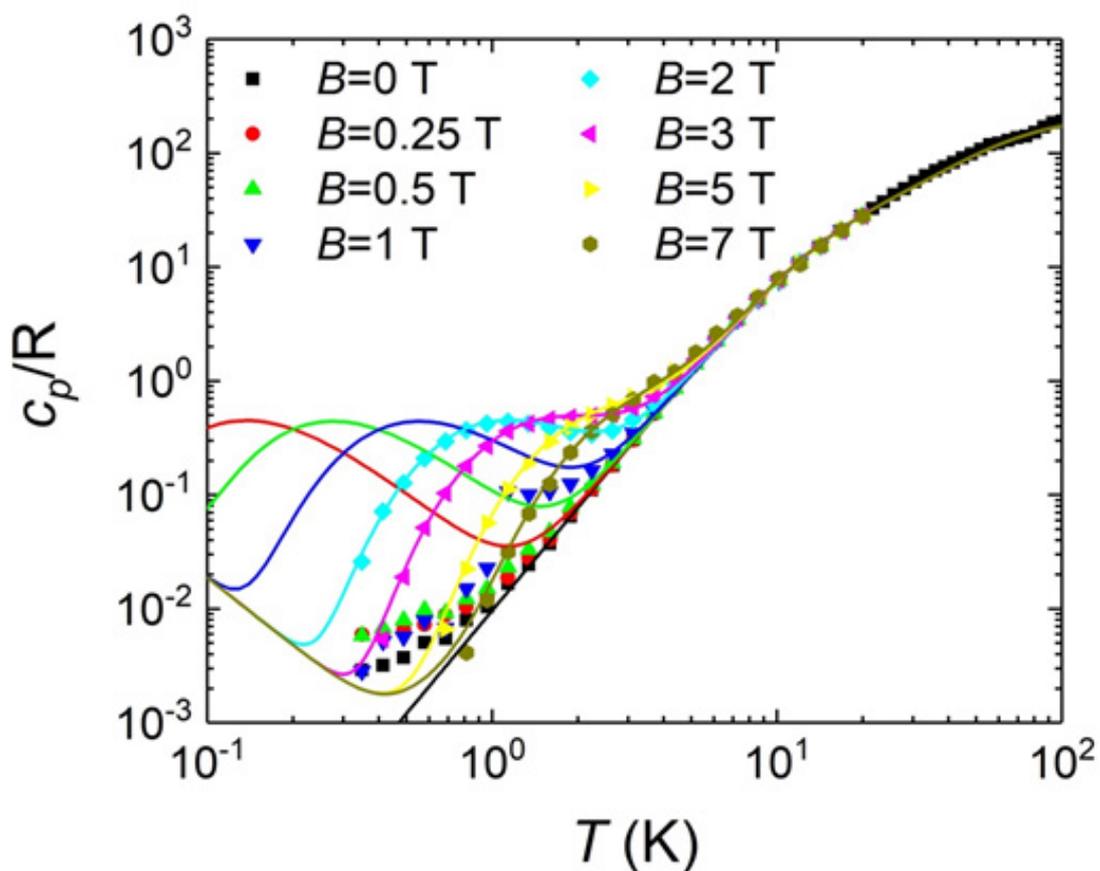

**Figure S24.** Heat capacity of a single crystal of **1**$^{VO}$ measured under different magnetic fields. Lines are simulations based on the parameters of the spin Hamiltonian derived from EPR measurements (see Fig. 2 of main text).

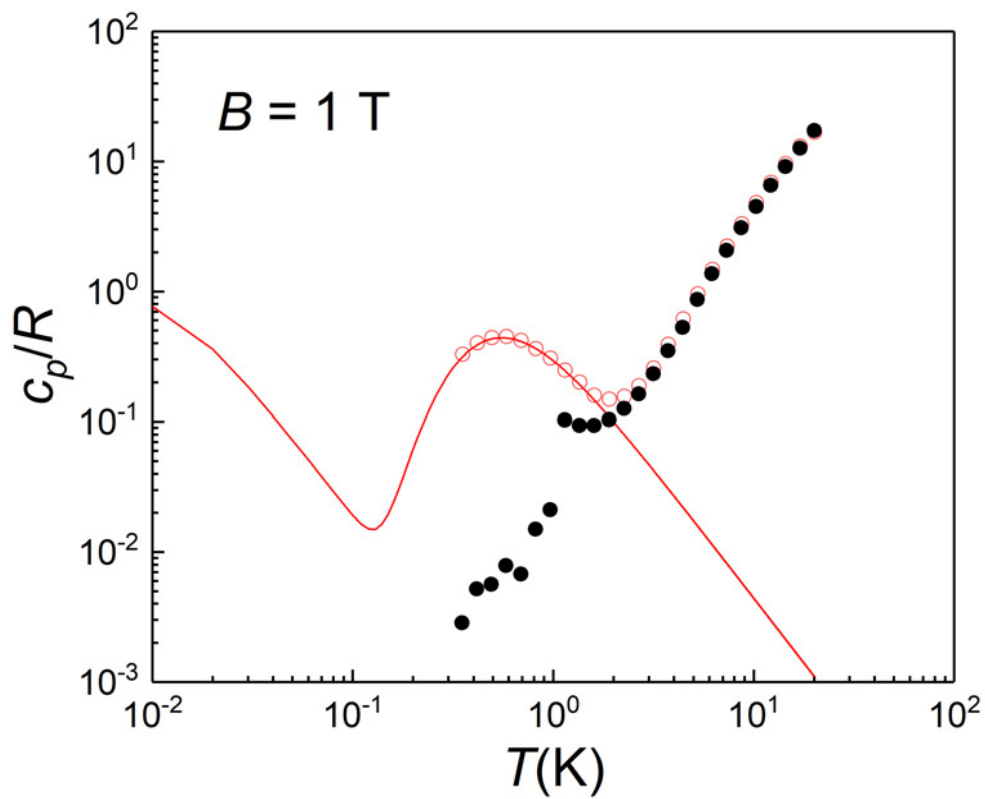

**Figure S25.** Specific heat of **1<sup>VO</sup>** measured at $B = 1$ T and for two different experimental times. At $T = 0.96$ K, the experimental time was 0.11 s for the solid dots and 4.4 s for the open symbols. In the former case, the spins are unable to attain equilibrium with the crystal lattice and the specific heat cannot reach its thermal equilibrium value, shown by the red solid line.

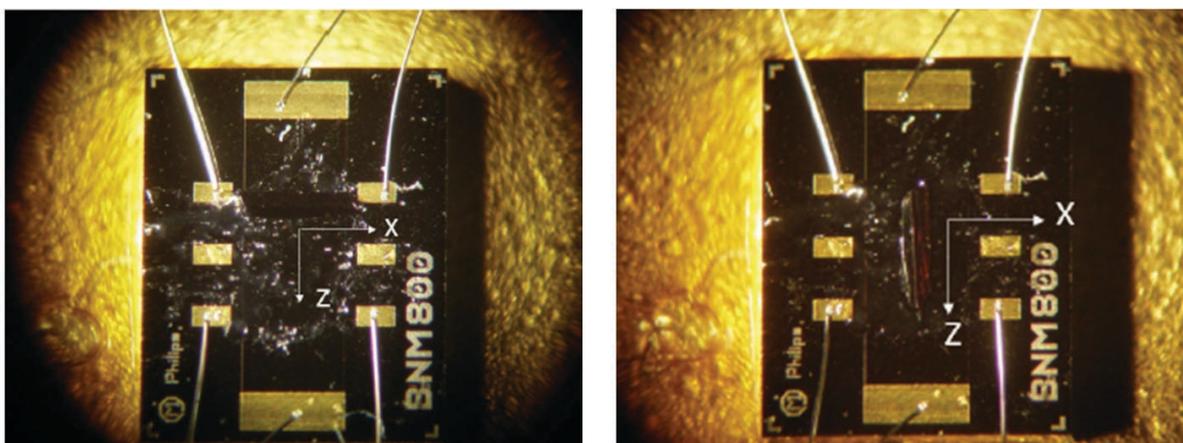

**Figure S26.** Pictures of a micro-Hall magnetometer hosting a single crystal of **1^VO** in two different orientations. The magnetic field was applied along the Z laboratory axis.

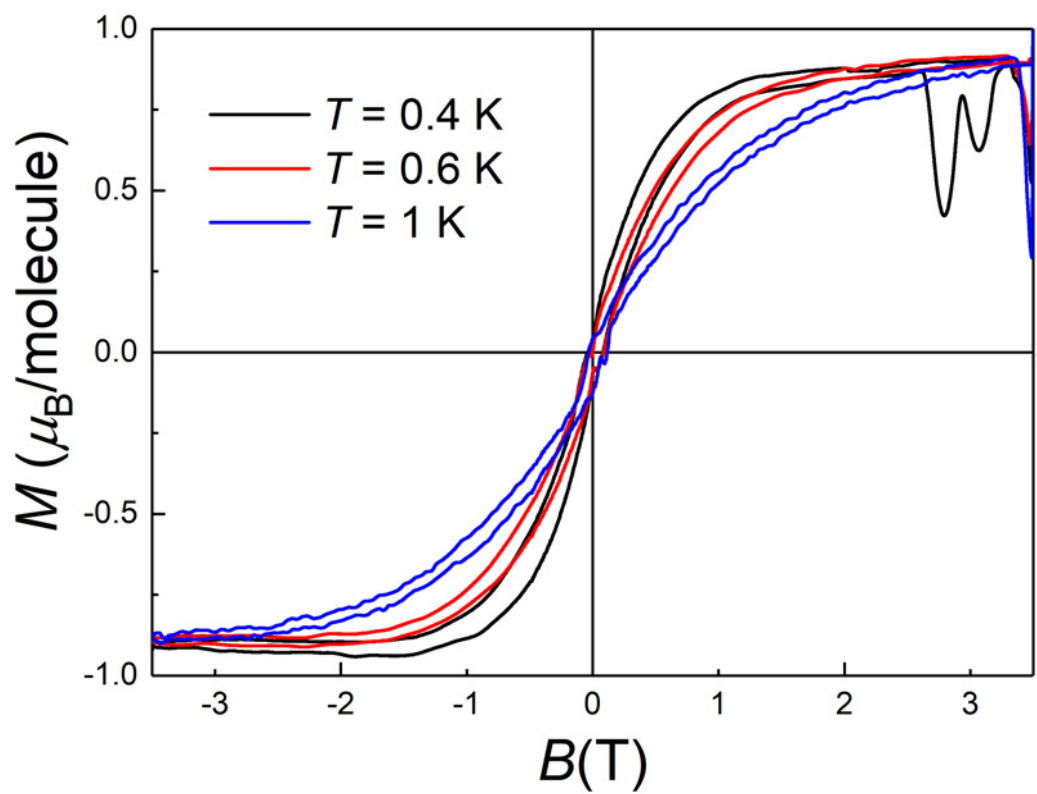

**Figure S27.** Magnetization hysteresis loops of a single crystal of **1^VO** measured with its long axis oriented along the X laboratory axis at different temperatures (left panel of Figure S26). The magnetic field sweeping rate was 1 T/min.

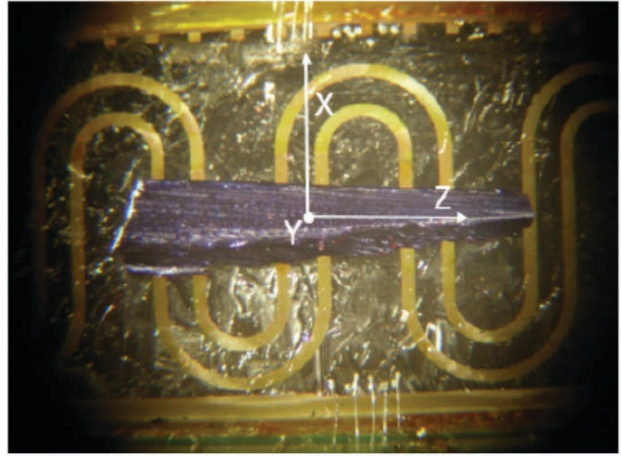

**Figure S28.** Picture of a Nb superconducting transmission line hosting a single crystal of **1**$^{VO}$. The magnetic field was applied along the X, Y and Z laboratory axis.

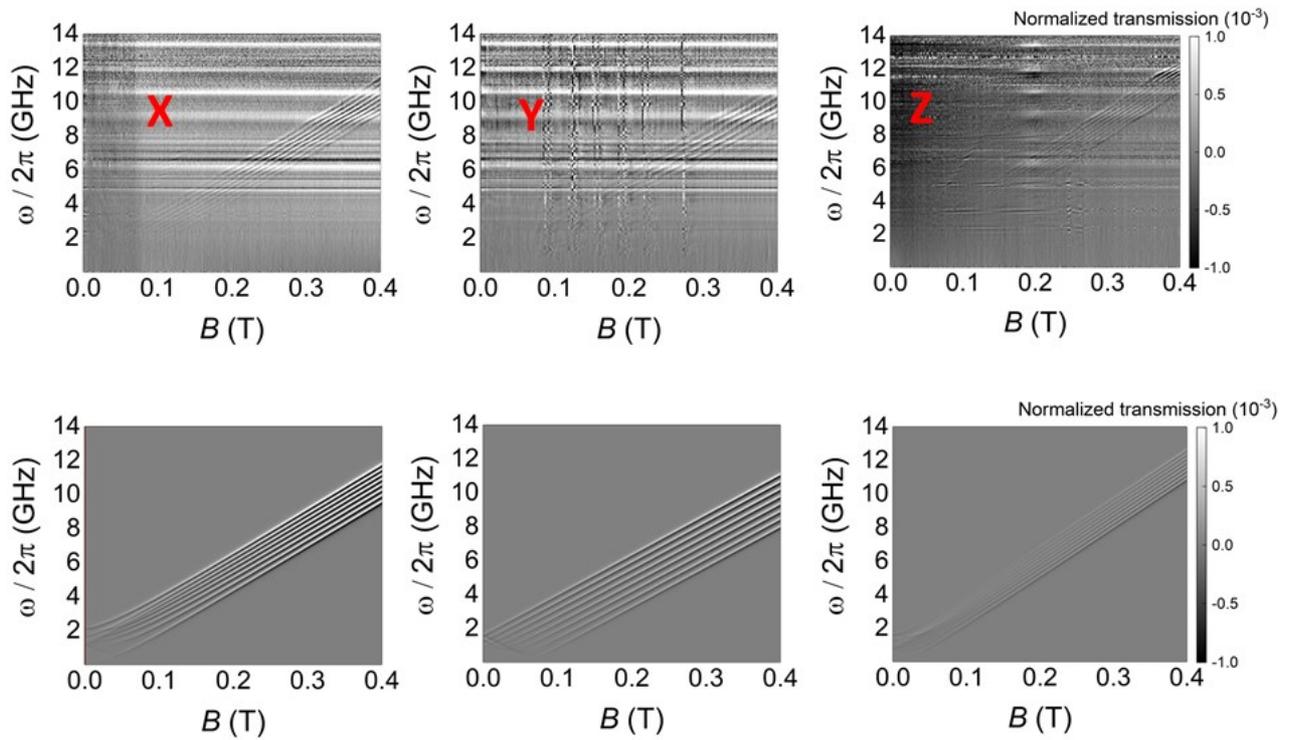

**Figure S29.** Grey scale two-dimensional plots of the normalized transmission through a superconducting transmission line coupled to a single crystal of **1**$^{VO}$ measured versus frequency ω and magnetic field applied along the laboratory axes X (left), Y (center) and Z (right). Top panels show experimental data and bottom panels show numerical simulations based on the spin Hamiltonian (Eq. (1) of the main text).

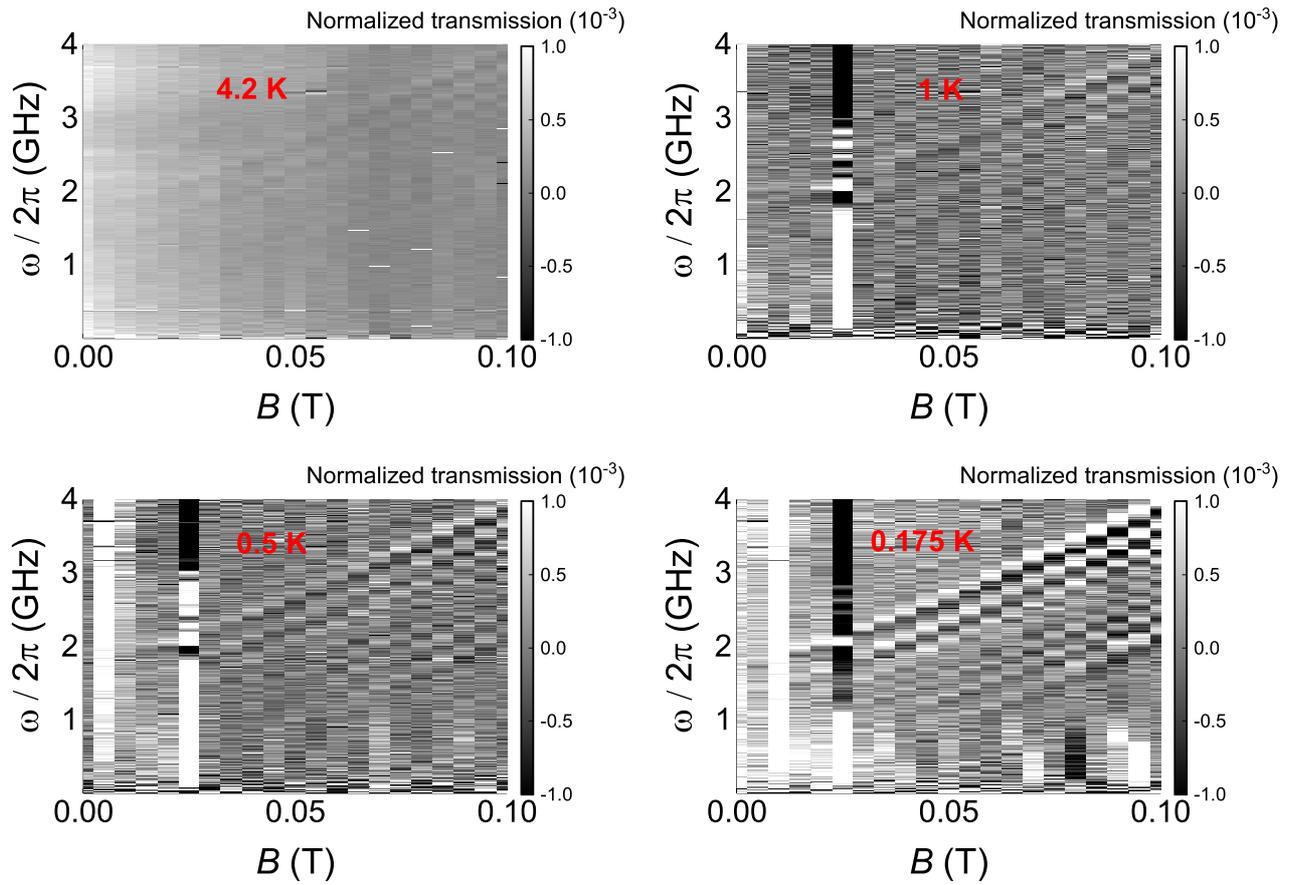

**Figure S30.** Low fields 2D plots of the normalized transmission through a superconducting transmission line coupled to a single crystal of **1^VO** measured versus frequency ω at decreasing temperatures as indicated. The magnetic field was applied along the Z laboratory axis.

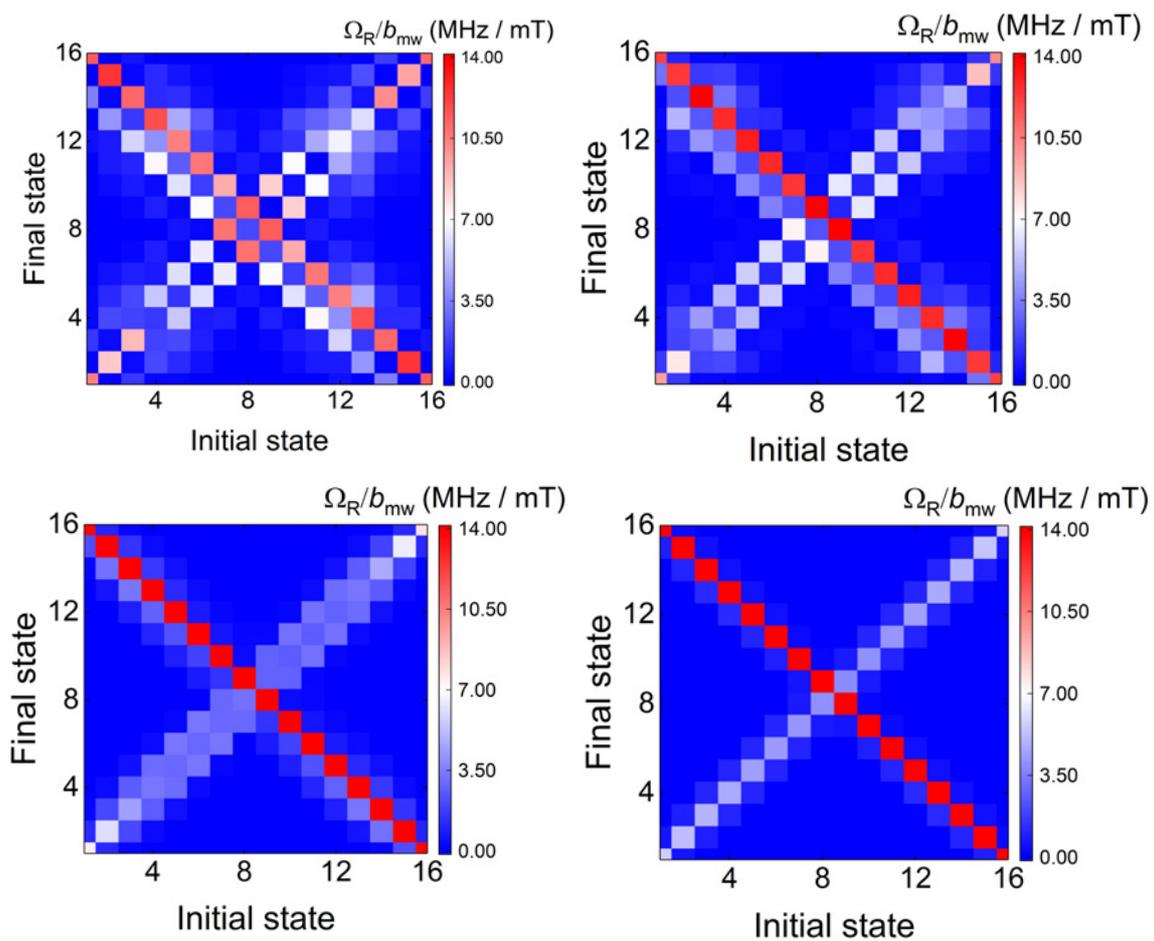

**Figure S31.** Plots of the Rabi frequencies between different electronuclear spin states of [VO(TCPPEt)], calculated from the spin Hamiltonian (Eq. (1) of the main text) and for progressively stronger (from top left to bottom right) magnetic field values $B$ = 0.02 T, 0.04 T, 0.1 T and 0.3 T. The field is applied along the X laboratory axis of Fig. S28.

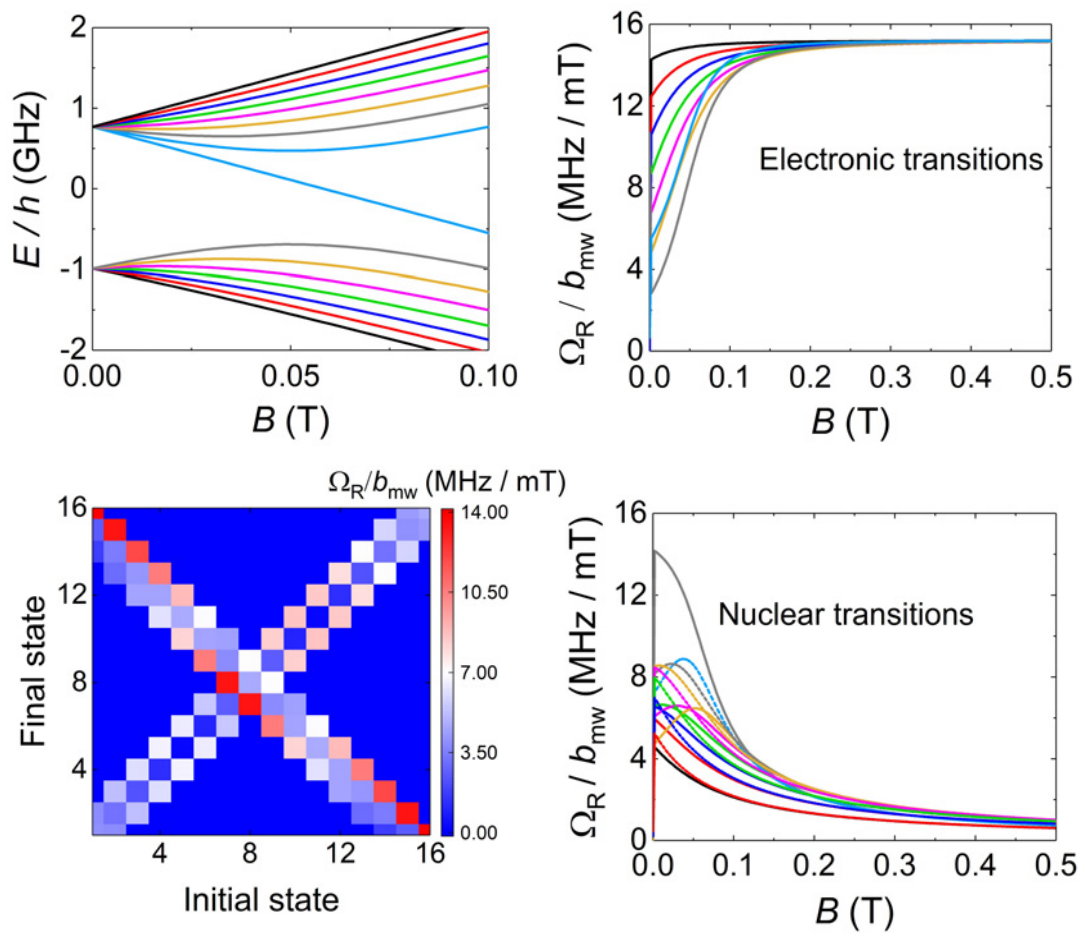

**Figure S32.** Left, top: energy levels of a $S = ½$ electronic spin coupled to a $I = 7/2$ nuclear spin coupled by an isotropic hyperfine interaction with $A_{\parallel} = A_{\perp} = 475$ MHz as a function of magnetic field applied along the X laboratory axis. Left, bottom: colour map of Rabi frequencies for resonant transitions, induced by a microwave magnetic field $b_{mw}$ applied along X, linking different electronuclear spin states at $B_X = 0.02$ T. Right: Magnetic field dependence of the Rabi frequencies of "electronic" (top) and "nuclear" (bottom) spin transitions.

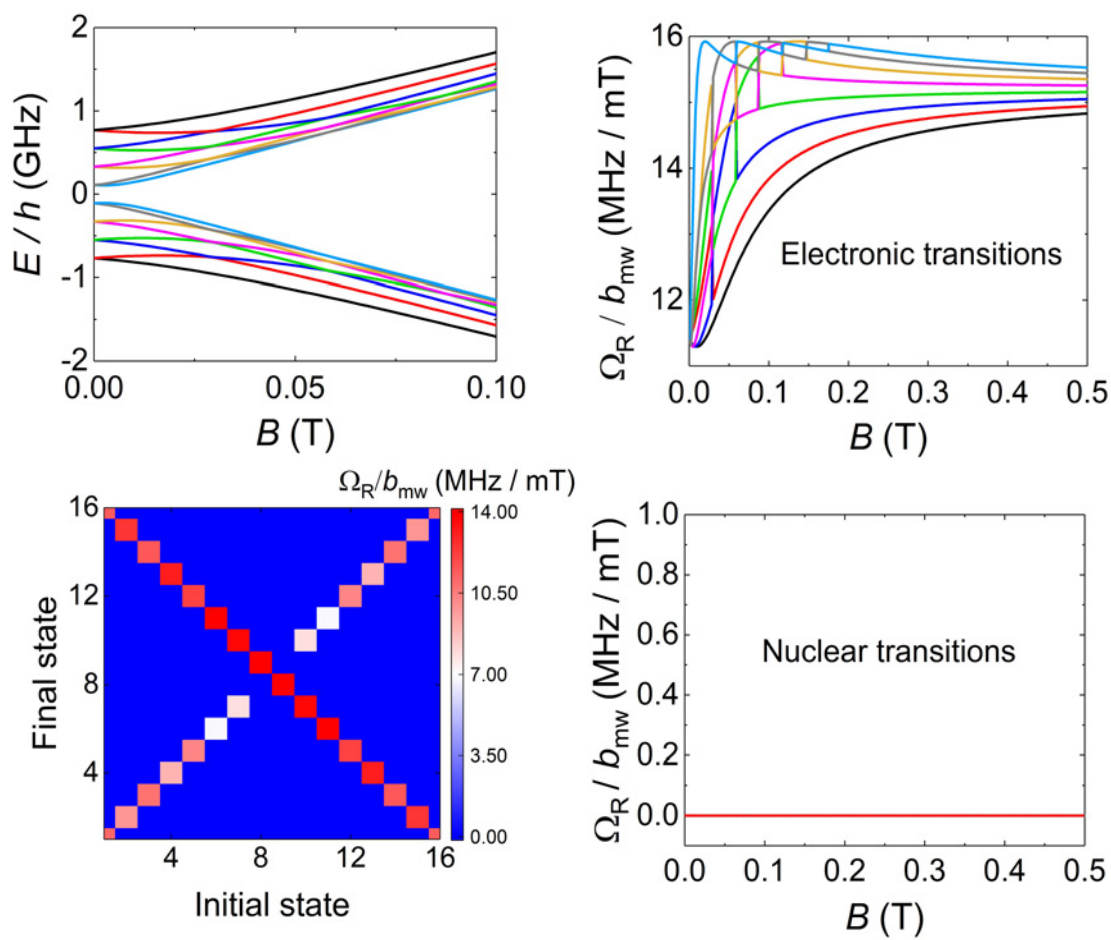

**Figure S33.** Left, top: energy levels of a $S = ½$ electronic spin coupled to a $I = 7/2$ nuclear spin by a fully uniaxial hyperfine interaction with $A_{||}$ = 475 MHz and $A_\perp = 0$ as a function of magnetic field applied along the X laboratory axis. Left, bottom: colour map of Rabi frequencies for resonant transitions, induced by a microwave magnetic field $b_{mw}$ applied along X, linking different electronuclear spin states at $B_X$ = 0.02 T. Right: Magnetic field dependence of the Rabi frequencies of "electronic" (top) and "nuclear" (bottom) spin transitions.

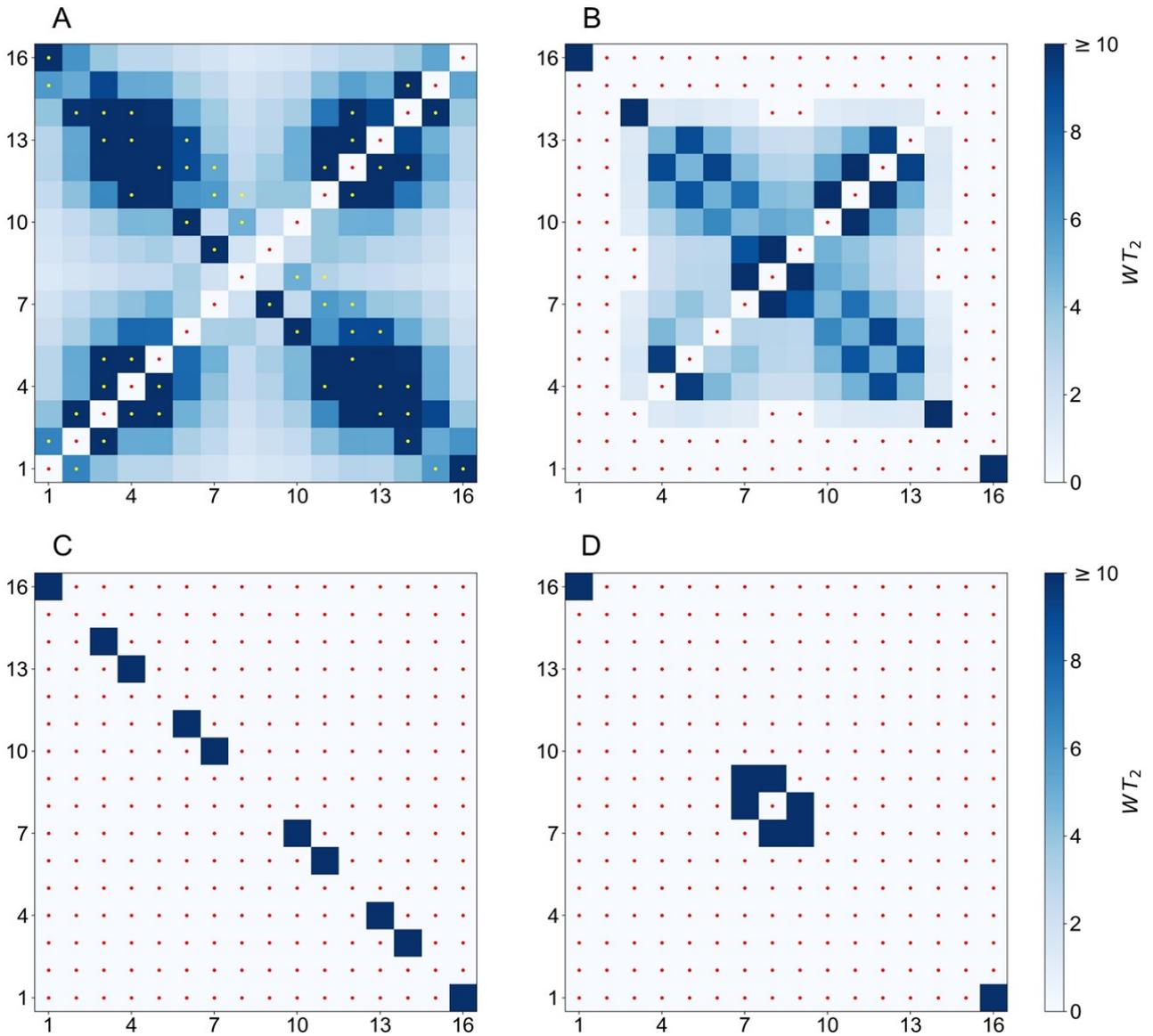

**Figure S34.** Universality plots, showing a 2D map of the operation rates linking different states, of a $S = ½$ and $I = 7/2$ electronuclear spin qudit ($d = 16$). The amplitude of resonant electromagnetic pulses was set to $b_{mw} = 1$ mT and $T_2 = 5$ μs. Yellow dots show the set of resonant transitions that are used to generate all operations. White spots, marked with a red dot at their center, signal pairs of states that cannot be connected by any sequence of such transitions. Panels A and B: calculations performed using the spin Hamiltonian and parameters of **1$^{VO}$** for, respectively, $B_X = 0.04$ and $0.1$ T. Panels C and D: calculations performed for a fully uniaxial ($A_\parallel = 475$ MHz and $A_\perp = 0$) and a fully isotropic ($A_\parallel = A_\perp = 475$ MHz) hyperfine interaction, respectively.

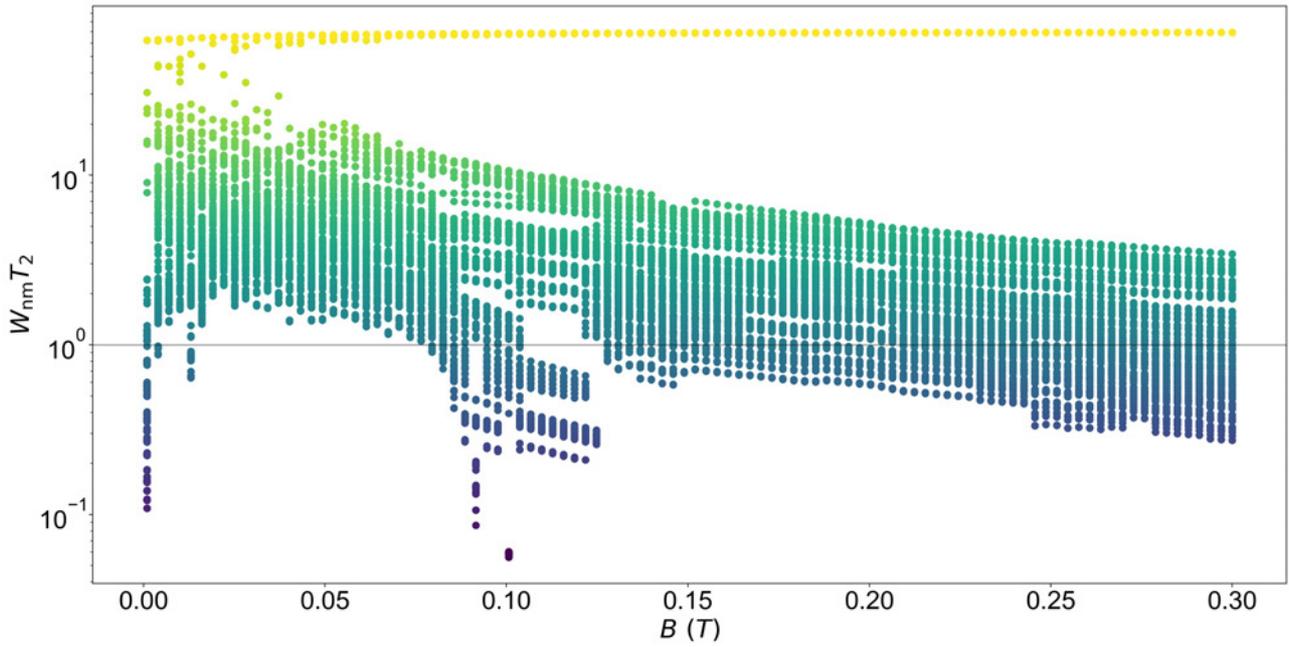

**Figure S35.** Operation rates $W_{nm}T_2$ between any pair of states of **1$^{VO}$** as a function of $B_X$, showing that more of them become effectively disconnected (for $W_{nm}T_2 < 1$) with increasing magnetic field. This means that the set of available quantum operations shrinks, thus the system cannot perform universal computations.